\documentclass[onecolumn,showpacs,preprintnumbers,elsart]{revtex4}
\usepackage{makeidx}
\usepackage{amssymb}
\usepackage{amsmath}
\usepackage{mathrsfs}
\usepackage{graphicx}
\usepackage{dcolumn}
\usepackage{bm}
\usepackage[center]{subfigure}
\usepackage{color}

\begin{document}

\title{Symmetry breaking in dipolar matter-wave solitons in dual-core
couplers}
\author{Yongyao Li$^{1,2,3}$, Jingfeng Liu$^{2}$, Wei Pang$^{4}$, and Boris
A. Malomed$^{1}$}
\email{malomed@post.tau.ac.il}
\affiliation{$^{1}$Department of Physical Electronics, School of Electrical Engineering,
Faculty of Engineering, Tel Aviv University, Tel Aviv 69978, Israel\\
$^{2}$Department of Applied Physics, South China Agricultural University,
Guangzhou 510642, China \\
$^{3}$Modern Educational Technology Center, South China Agricultural
University, Guangzhou 510642, China\\
$^{4}$ Department of Experiment Teaching, Guangdong University of
Technology, Guangzhou 510006, China.}

\begin{abstract}
We study effects of the spontaneous symmetry-breaking (SSB) in solitons
built of the dipolar Bose-Einstein condensate (BEC), trapped in a dual-core
system with the dipole-dipole interactions (DDIs) and hopping between the
cores. Two realizations of such a matter-wave coupler are introduced,
weakly- and strongly-coupled. The former one in based on two parallel
pipe-shaped traps, while the latter one is represented by a single pipe
sliced by an external field into parallel layers. The dipoles are oriented
along axes of the pipes. In these systems, the dual-core solitons feature
the SSB of the supercritical type and subcritical types, respectively.
Stability regions are identified for symmetric and asymmetric solitons, and,
in addition, for non-bifurcating antisymmetric ones, as well as for
symmetric flat states, which may also be stable in the strongly-coupled
system, due to competition between the attractive and repulsive intra- and
inter-core DDIs. Effects of the contact interactions are considered too.
Collisions between moving asymmetric solitons in the weakly-symmetric system
feature elastic rebound, merger into a single breather, and passage
accompanied by excitation of intrinsic vibrations of the solitons, for
small, intermediate, and large collision velocities, respectively. A $%
\mathcal{PT}$-symmetric version of the weakly-coupled system is briefly
considered too, which may be relevant for matter-wave lasers. Stability
boundaries for $\mathcal{PT}$-symmetric and antisymmetric solitons are
identified.
\end{abstract}

\maketitle



\section{Introduction}

Studies of Bose-Einstein condensates (BECs) made of dipolar atoms or
molecules had produced a great deal of fascinating experimental and
theoretical results, which were summarized in recent reviews \cite%
{manual,Lahaye} and \cite{Baranov,Lahaye}, respectively. The continuation of
the work in this direction has yielded new remarkable findings, such as the
prediction of various pattern-formation mechanisms \cite%
{Patterns,roton-layered} (which share some features with the formation of
patterns in ferrofluids \cite{Richter}), analysis of the stability of the
dipolar BEC trapped in optical-lattice (OL) potentials \cite{OL} and of the
roton instability \cite{roton-layered,roton}, the possibility of the
Einstein - de Haas effect \cite{Haas}, etc. Important experimental
achievements, which offer new perspectives for studies of dipole-dipole
interactions (DDIs) in atomic condensates, are the creation of BEC in
dysprosium \cite{Dy} and erbium \cite{Er}. Parallel to that, essential
results have been obtained for degenerate quantum gases of dipolar fermions
\cite{Baranov,Fermi}.

In addition to their own physical significance, dipolar condensates may also
be used as \textit{quantum simulators} \cite{simulator,Lewenstein}
representing other physical media where nonlocal nonlinearities play a
fundamental role. These include the heating and ionization of plasmas \cite%
{Litvak}, nonlinear optics of nematic liquid crystals \cite{Peccianti}, of
waveguides sensitive to temperature variations \cite{Krolik}, and of
semiconductor cavities \cite{Falk}, BECs with long-range interactions
induced by laser illumination \cite{Gershon}, and others settings.

An interesting ramification of the study of collective nonlinear modes in
the dipolar BEC is the prediction of solitons (which have not yet been
reported in experimental works). In effectively one-dimensional (1D) traps,
solitons were analyzed in both continual \cite{1D-cont} and discrete \cite%
{1D-discr} settings, the latter one corresponding to the fragmentation of
the BEC by a deep OL. In a similar form, 1D solitons supported by the
attractive DDIs were predicted in the Tonks-Girardeau gas of dipolar
hard-core bosons \cite{TG}. Taking into account the 3D structure of the
quasi-1D cigar-shaped traps, the solitons, including ones with embedded
vorticity (cf. similar modes introduced earlier in the context of BEC\ with
local interactions \cite{Luca}) were further studied in Ref. \cite{SKA1},
and gap solitons in a similar setting, but including an OL potential, were
considered too \cite{SKA2}. In the 2D system, discrete fundamental solitons
and solitary vortices with long-range DDIs between sites of the lattice can
be constructed easily \cite{2D-discr}. In the continual 2D model,
fundamental \cite{Pedri} and vortical \cite{vortex} solitons were
constructed in the isotropic setting, assuming that the sign of the DDI
could be reversed from repulsion to attraction by means of a rapidly
oscillating ac field \cite{inversion}. 2D fundamental and vortex solitons
supported by a trapping potential were introduced in Ref. \cite{Lashkin}.

Without reversing the DDI sign, stable 2D anisotropic solitons,
corresponding to the in-plane polarization of dipoles, were constructed in
Ref. \cite{Tikhonenkov}, by means of the variational approximation and
systematical numerical simulations. The variational approximation for 2D
solitons supported by the DDI was analyzed in Ref. \cite{Wunner-VA}, and a
rigorous proof of the existence of such solitons was provided too \cite%
{rigorous}. Also studied were more complex situations, such as the formation
of a multi-soliton patterns as a result of the development of the
modulational instability of an extended state \cite{MI-formation}.

The long-range character of the DDI makes it possible to consider
interactions between condensate layers trapped in parallel planar
waveguides. The DDI couples them by nonlinear forces even in the absence of
hopping (tunneling) of atoms across gaps separating the layers \cite%
{roton-layered} (the isolation of parallel layers can be provided by a
strong OL field whose axis is perpendicular to the layers \cite{OL-first}).
This nonlocal interaction gives rise to ``indirect" scattering of 2D
solitons moving in the separated layers \cite{stack-scattering}, and the
formation of bound states of such solitons \cite{stack-molecule}. The
creation of multi-soliton filaments and checkerboard crystals in
multi-layered stack was predicted too \cite{crystal-in-stack}.

The model of nonlocal DDIs between parallel layers considered in Refs. \cite%
{stack-scattering,roton-layered,crystal-in-stack,stack-molecule} did not
take into regard the hopping (tunneling of atoms, alias \textit{linear
coupling}) between the layers. On the other hand, models of \textit{%
dual-core couplers}, with intrinsic local nonlinearity acting in both cores,
were studied in detail in terms of optics and matter waves, starting from
the analysis of the spontaneous symmetry breaking (SSB) of CW
(continuous-wave, i.e., uniform) states in dual-core optical fibers with the
cubic and more general forms of the intra-core nonlinearity. In that system,
the linear coupling is caused by the overlap of the evanescent field,
originating from each core, with the parallel one. The SSB happens in the
dual-core fiber, as a result of the interplay of the linear coupling and
intrinsic nonlinearity, with the increase of the total power of the CW beam.
The analysis of the SSB was extended, in full detail, to temporal and
spatial solitons \cite{fibers,fibers2}, and to optical domain walls \cite{DW}%
. The SSB effects were also studied for solitons in dual-core fiber Bragg
gratings \cite{BraggSSB}, in two-tier waveguiding arrays (for discrete
solitons) \cite{Amherst}, in parallel-coupled waveguides with the quadratic
(second-harmonic-generating) \cite{chi2} and cubic-quintic (CQ) \cite{CQ}
nonlinearities, as well as for dissipative solitons in linearly-coupled CQ
complex Ginzburg-Landau equations \cite{CGL}. Recently, a similar analysis
was developed for the SSB of solitons in $\mathcal{PT}$-symmetric couplers,
with mutually balanced loss and gain (and identical cubic nonlinearities)
acting in the two cores \cite{PT,PT2}. Unlike the above-mentioned settings,
in the latter case the SSB destroys symmetric solitons, rather than
replacing them by stable asymmetric ones. The linear coupling in optics may
also represent the mutual interconversion of two polarizations of light in
twisted fibers (which, in particular, are used in the so-called rocking
filters) \cite{twist}, twisted photonic-crystal fibers \cite{PCF-twist},
twisted fiber gratings \cite{grating-twist}, or the interconversion of two
waves with different carrier frequencies, caused by the
electromagnetically-induced transparency \cite{yongyao}.

Similar dual-core (alias \textit{double-well}) settings, approximated by
linearly-coupled Gross-Pitaevskii equations (GPEs), were introduced for the
mean-field wave functions describing trapped BEC with local interactions
\cite{Arik}. A similar linear coupling accounts for the mutual
interconversion in a mixture of two different atomic states, induced by a
resonant electromagnetic wave \cite{Ballagh}. The linearly coupled GPEs were
used to predict the shift of the miscibility-immiscibility transition in BEC
or fermionic mixtures of two states connected by the linear interconversion
\cite{Merhasin} and the stabilization of 2D solitons against the collapse in
a linearly-coupled binary system with attractive and repulsive
intra-component interactions \cite{Ueda}. The SSB of matter-wave solitons
trapped in 1D and 2D linearly-coupled cores was analyzed in Ref. \cite{Arik}%
. In the 1D situation, a more accurate description, which, nevertheless,
yields similar results for the solitons' SSB, is provided by the
two-dimensional GPE, which, instead of postulating two 1D wave functions in
the two parallel cores with the linear exchange between them, introduces a
single 2D wave function comprising both cores \cite{Marek}. This includes
the case when the cores are defined by nonlinear \textit{pseudopotentials}
(rather than by a linear trapping potential), i.e., local modulation of the
self-attraction coefficient \cite{Hung}.

Although the previous works analyzed many aspects of the SSB in dual-core CW
and solitonic states, those works were dealing solely with local intrinsic
nonlinearities. Only in a very recent paper \cite{Fangwei}, a shift of the
SSB transition of solitons in the coupler with nonlocal nonlinearity of the
thermal type, typical to optical systems \cite{Krolik}, was considered. The
analysis was performed for two opposite limit cases, \textit{viz}., the weak
nonlocality characterized by a small correlation radius, which may be
approximated by the first two terms of the expansion of the nonlocal cubic
term, and the opposite limit of the infinite correlation radius, which
corresponds to the two-component quasi-linear model of ``accessible
solitons" \cite{accessible}.

The aim of the present work is to consider the SSB of solitons in the
effectively 1D dual-core coupler filled by the dipolar condensate, which
exhibits the interplay of the long-range DDIs and linear hopping between the
quasi-1D cores. In particular, the DDIs act both inside the cores and
between them, while the thermal nonlocality considered in Ref. \cite{Fangwei}
could not act across the gap separating the parallel waveguides. It is
relevant to stress that, in the absence of the longitudinal dimension, the
double-well setting is not sufficient to exhibit the nonlocal character of
the interactions in the dipolar condensate, the minimum necessary
configuration being based on a set of three potential wells \cite{triple}.

\begin{figure}[tbp]
\centering\subfigure[] {\label{fig1a}
\includegraphics[scale=0.4]{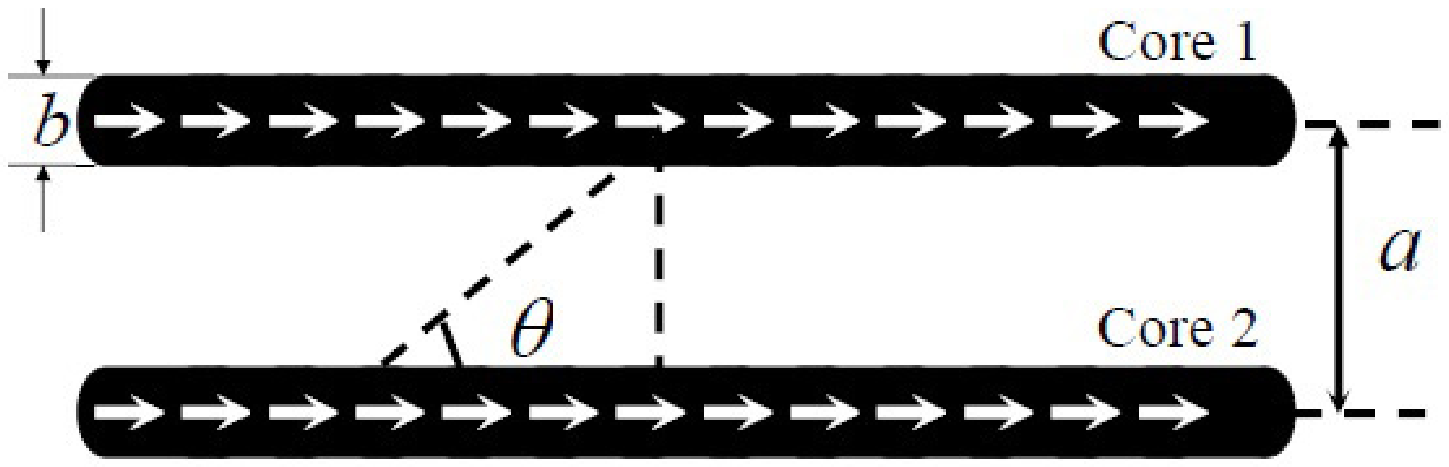}}%
\subfigure[] {\label{fig1b}
\includegraphics[scale=0.35]{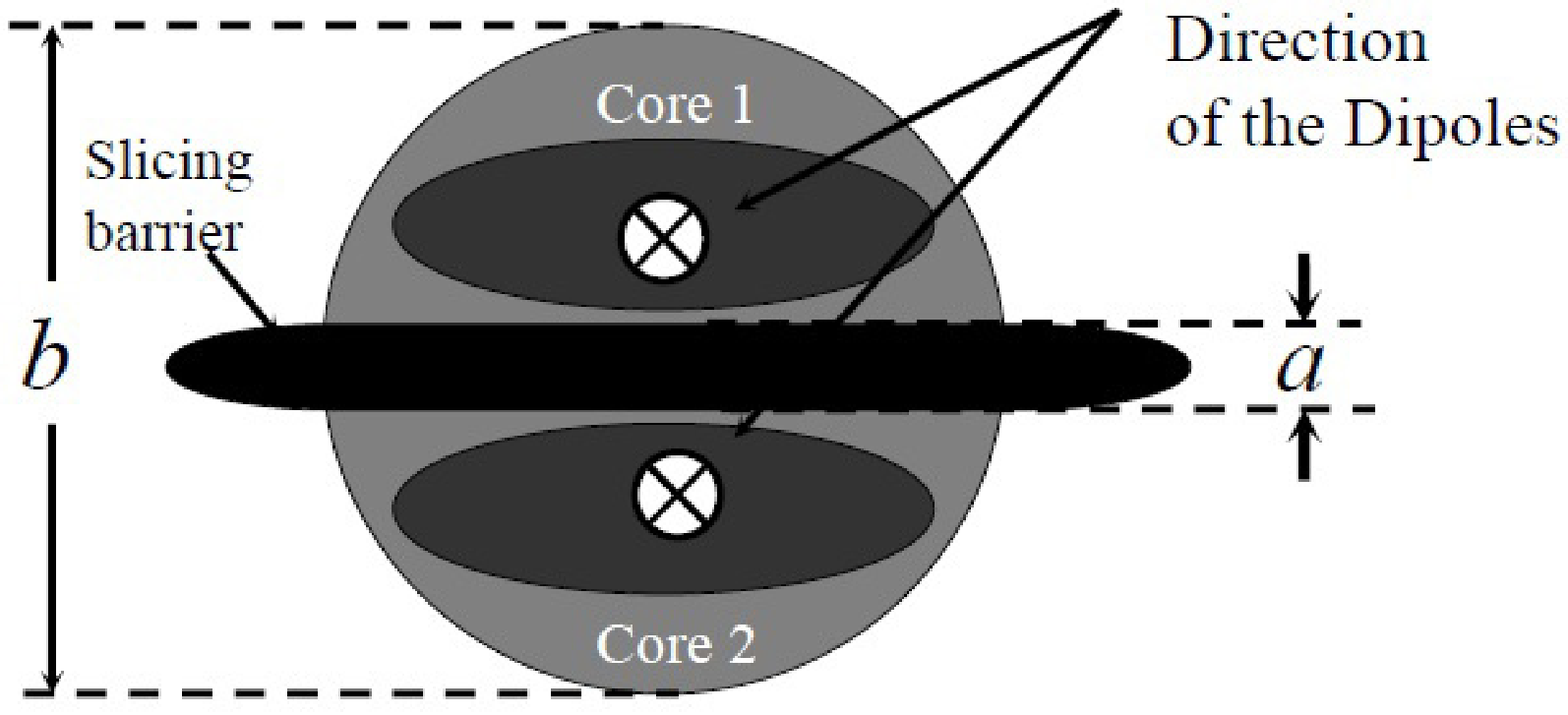}}
\caption{(Color online) (a) In the \textit{weakly-coupled system}, the
dipolar condensates are trapped in parallel pipes of diameter $b$, separated
by distance $a$. The arrows represent the orientation of the dipoles. (b)
The \textit{strongly-coupled system} is shown by means of the cross-section
image of the condensate trapped in the single pipe of diameter $b$, which is
sliced by a repelling laser sheet into two layers, with effective separation
$a$ between them. Symbol $\otimes $ represents the orientation of the
dipoles, which are perpendicular to the figure's plane. }
\label{Fig1}
\end{figure}

Two coupler configurations are considered here, as shown in Fig. \ref{Fig1}.
The first setting, presented in Fig. \ref{fig1a}, is based on two identical
condensate-trapping pipes of diameter $b$, separated by distance $a$. In
this case, $b<a$ is implied, hence the system may be naturally called a
\textit{weakly-coupled} one. The other setting is shown in Fig. \ref{fig2b},
with the condensate loaded into the single pipe of diameter $b$, which is
sliced into two parallel layers by a thin potential barrier of small
thickness $a$. The barrier can be induced by a repulsive light sheet
(blue-shifted one, with respect to the atoms) \cite{sheet}. The latter
setting implies $b>a$, and it will be named, accordingly, a \textit{%
strongly-coupled} system. In either case, the dipoles are polarized along
the pipes' axes, hence the DDIs are attractive inside the cores, which makes
it possible to form solitons in each one \cite{1D-cont}.

Aiming to study the SSB, alias the symmetry-breaking phase transition, in
these settings, it is relevant to recall that there are two types of the SSB
\textit{bifurcation}, namely, the subcritical and supercritical ones \cite%
{Joseph}, which are tantamount to the phase transitions of the first and
second kinds, respectively. In the subcritical situation, branches
representing asymmetric modes emerge as unstable states at the bifurcation
point, then go backward in the bifurcation diagram, and get stabilized after
turning forward. In the supercritical setting, the asymmetric branches
emerge as stable ones at the bifurcation point and immediately continue
forward.

The rest of the paper is organized as follows. In Sec. II, we construct
symmetric solitons in the two settings presented in Fig. 1, and then
identify the SSB transitions to asymmetric solitons. Antisymmetric solitons
and their stability are considered too (these solitons do not undergo any
bifurcation). Basic results for the stability of different modes are
presented (including a flat state, which may also be stable in the
strongly-coupled system). Effects of the local (contact) nonlinearity on the
SSB are considered too. In Sec. III, we study collisions between moving
asymmetric solitons. In Sec. IV, a $\mathcal{PT}$-symmetric \cite{Bender}
extension of this system is briefly considered, which includes gain and loss
applied to the two cores (similar to the $\mathcal{PT}$-symmetric coupler
with the local nonlinearity introduced in Refs. \cite{PT,PT2}). The paper is
concluded by section V.

\section{The symmetry-breaking bifurcation and stability of solitons}

\subsection{The coupled Gross-Pitaevskii equations}

\ In the usual mean-field approximation \cite{Lahaye,Baranov}, both
dual-core settings introduced in Fig. \ref{Fig1} are described by the system
of 1D linearly-coupled GPEs for the wave functions in the two cores, $\psi
_{1}$ and $\psi _{2}$. In the scaled form, the equations are
\begin{gather}
i{\frac{\partial \psi _{n}}{\partial t}}=-{\frac{1}{2}}{\frac{\partial
^{2}\psi _{n}}{\partial x^{2}}}+g|\psi _{n}|^{2}\psi _{n}-\kappa \psi _{3-n}
\notag \\
-G_{\mathrm{DD}}\psi _{n}(x)\int_{-\infty }^{+\infty }\left[ {\frac{|\psi
_{n}(x^{\prime })|^{2}}{(b^{2}+|x-x^{\prime }|^{2})^{3/2}}}-\frac{1}{2}{%
\frac{(1-3\cos ^{2}\theta )|\psi _{3-n}(x^{\prime })|^{2}}{%
2(a^{2}+|x-x^{\prime }|^{2})^{3/2}}}\right] dx^{\prime },  \label{NLS}
\end{gather}%
where $n=1,2$ and $\cos \theta =|x-x^{\prime }|/(a^{2}+|x-x^{\prime
}|^{2})^{1/2}$, see Fig. \ref{Fig1}(a). In this notation, $\kappa $ is the
coupling parameter (hopping coefficient), $g$ represent the local
interaction (repulsive in the case of $g>0$), and the orientation of the
dipoles in Fig. \ref{Fig1} corresponds to $G_{\mathrm{DD}}>0$, which implies
the attraction between dipoles in the given core and repulsion between the
cores for $\cos ^{2}\theta <1/3$. The first term in the integrand, that
accounts for the intra-core DDIs, is regularized by the transverse diameter,
$b$, which is an approximation sufficient for producing 1D solitons \cite%
{1D-cont}.

The solitons will be characterized by the total norm (proportional to the
number of atoms in the condensate),
\begin{equation*}
P\equiv P_{1}+P_{2}=\int_{-\infty }^{+\infty }\left( |\psi _{1}|^{2}+|\psi
_{2}|^{2}\right) dx.
\end{equation*}%
To focus on SSB effects dominated by the DDI, we will first drop the local
nonlinearity, setting $g=0$ (in the experiment, this can be done by means of
the Feshbach resonance \cite{Feshbach}); effects of the contact interactions
will be considered afterwards. Then, we scale the units to fix $\kappa
\equiv 1$ and $G_{\mathrm{DD}}\equiv 1$, the remaining free parameters being
$a$, $b$, and $P$ (and $g$ too, in the end).

Stationary solutions to Eq. (\ref{NLS}) with chemical potential $\mu $ are
looked for in the usual form, $\psi _{1,2}(x,t)=\exp \left( -i\mu t\right)
\phi _{1,2}(x)$, with real functions $\phi _{1,2}(x)$. In particular, $\phi
_{1,2}(x)\equiv \phi (x)$ for symmetric solutions obeys the stationary
equation,%
\begin{gather}
\left( \mu +1\right) \phi =-{\frac{1}{2}}{\frac{d^{2}\phi }{dx^{2}}}+g\phi
^{3}  \notag \\
-\phi (x)\int_{-\infty }^{+\infty }\left[ {\frac{1}{(b^{2}+|x-x^{\prime
}|^{2})^{3/2}}}-{\frac{1-3\cos ^{2}\theta }{2(a^{2}+|x-x^{\prime
}|^{2})^{3/2}}}\right] \phi ^{2}(x^{\prime })dx^{\prime },  \label{phi}
\end{gather}%
where $\kappa =G_{\mathrm{DD}}=1$ is fixed, as said above. Equations (\ref%
{NLS}) and (\ref{phi}) are solved below by means of numerical methods. In
particular, stationary solutions were found below by means of the
imaginary-time propagation method \cite{imaginary} with periodic boundary
condition, while the stability of the solutions was tested by means of the
integration in real time.

\subsection{The weakly-coupled system ($a>b$)}

\begin{figure}[tbp]
\centering\subfigure[] {\label{fig2a}
\includegraphics[scale=0.3]{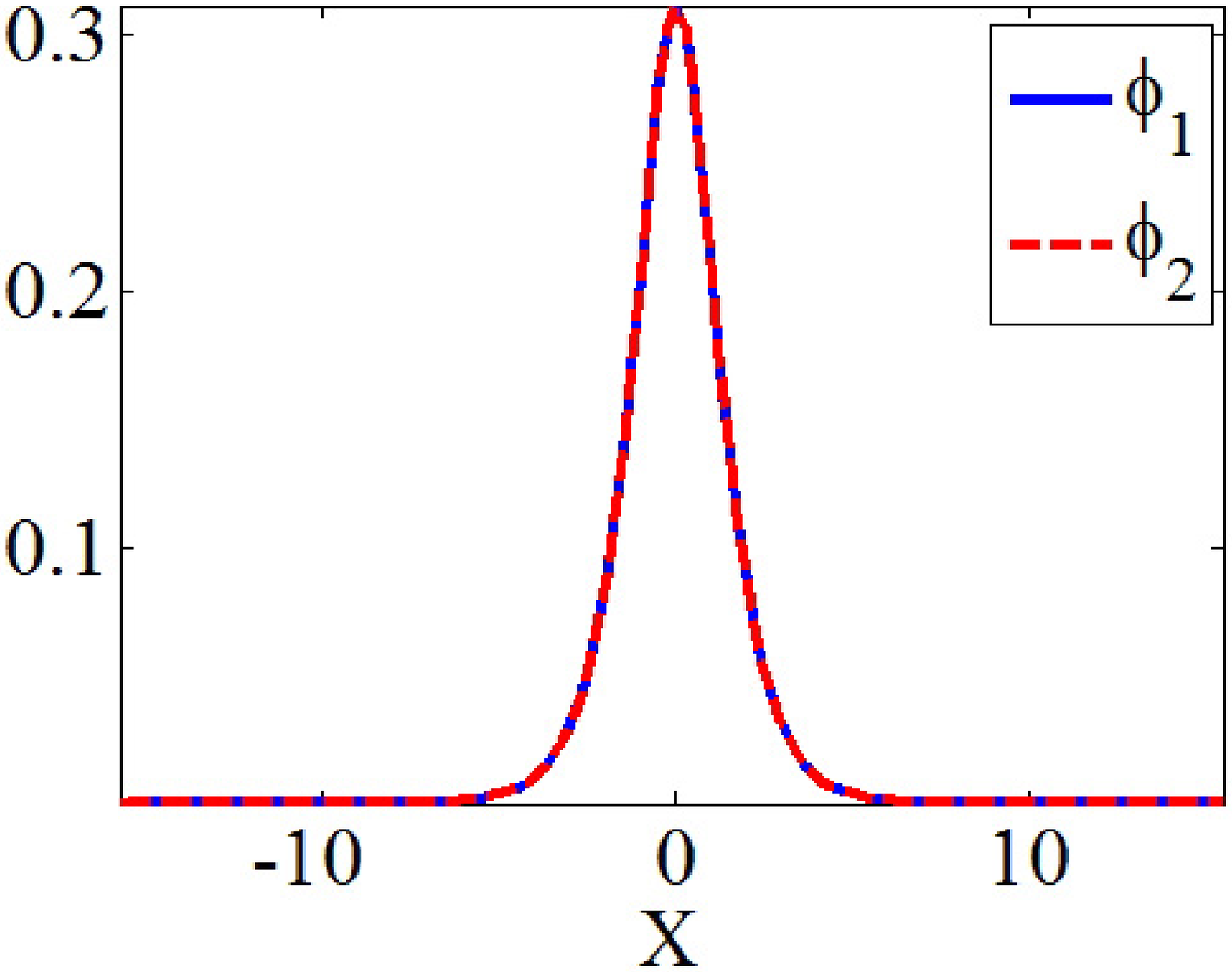}}%
\subfigure[] {\label{fig2b}
\includegraphics[scale=0.3]{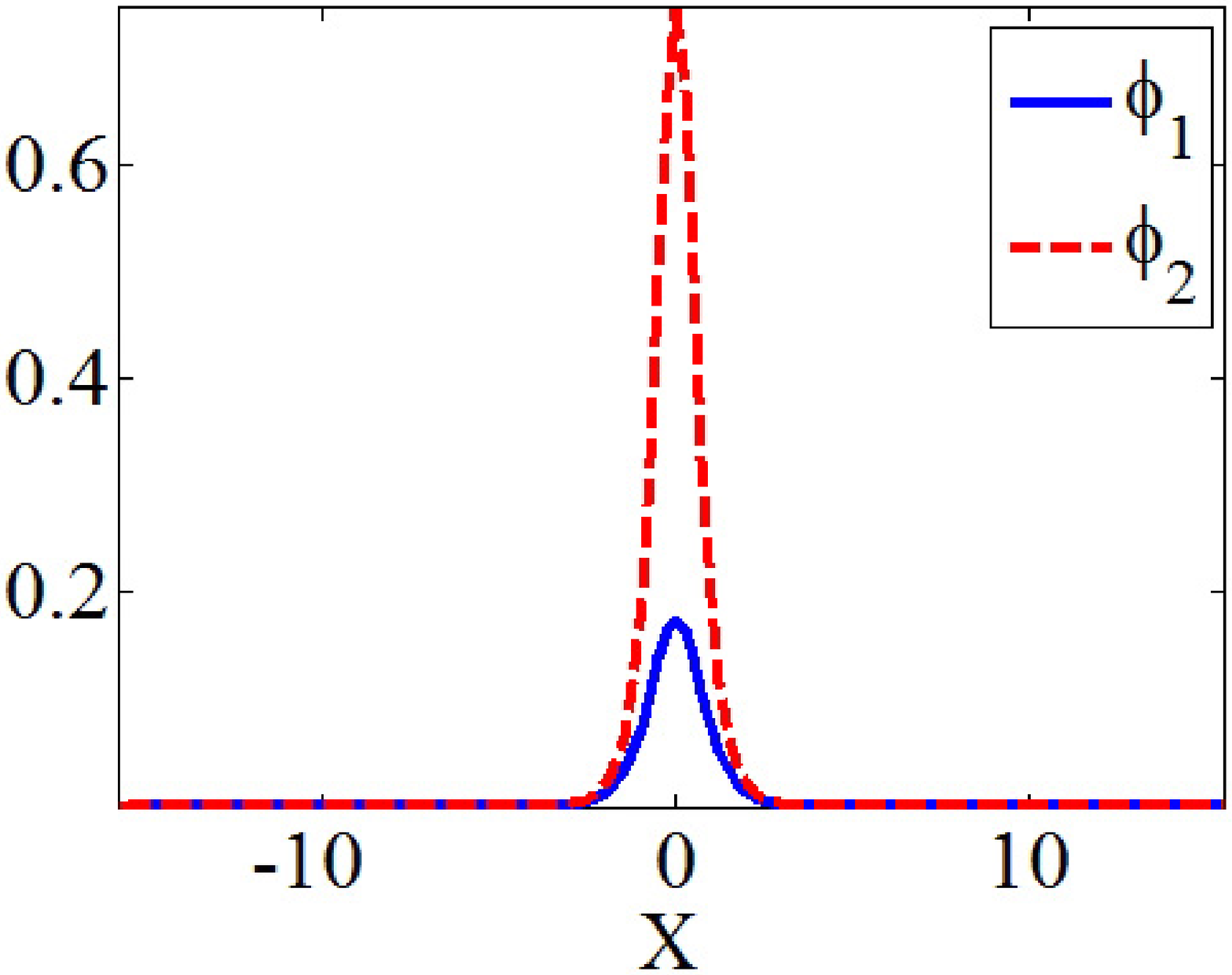}}
\subfigure[] {\label{fig2c}
\includegraphics[scale=0.3]{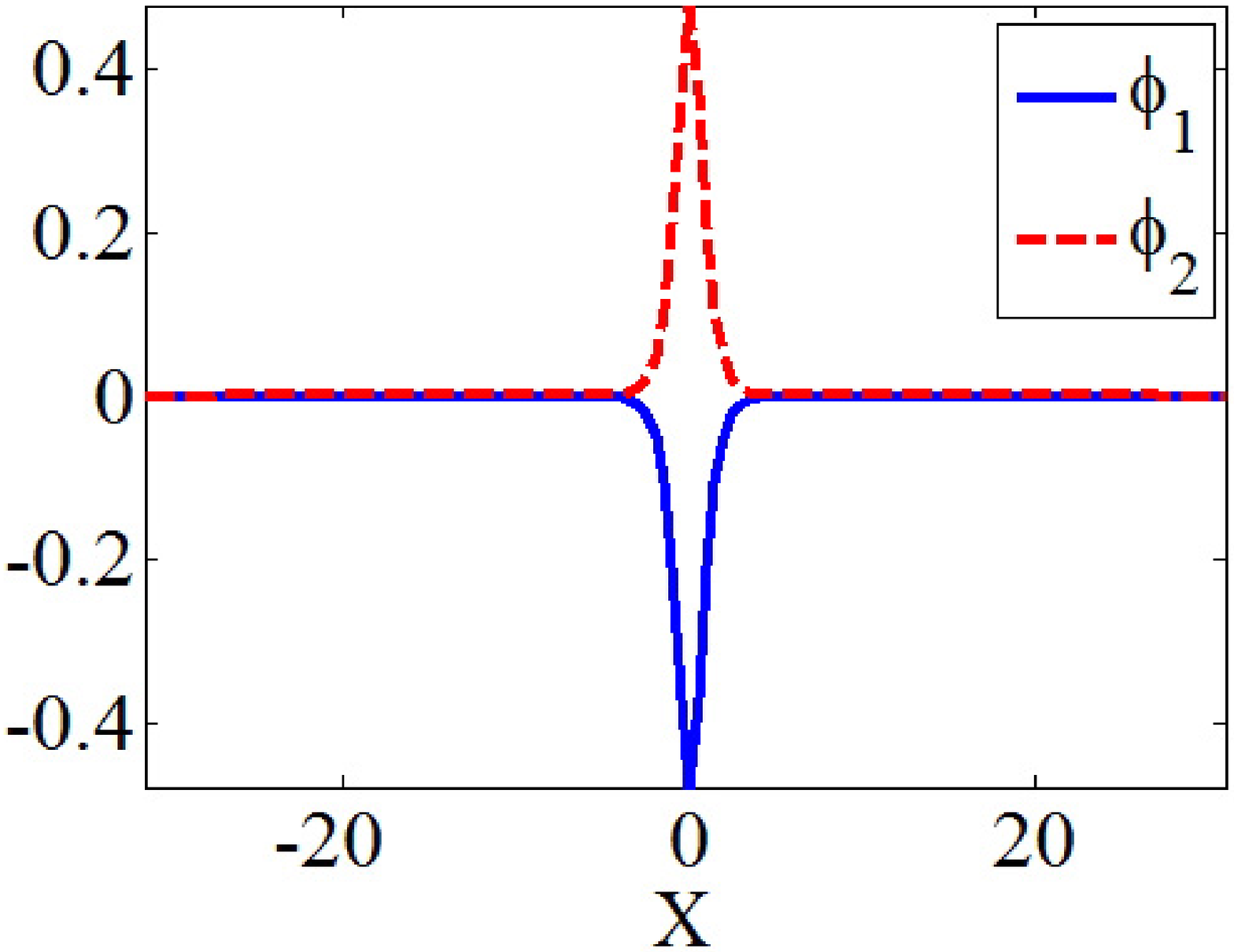}}
\subfigure[] {\label{fig2d}
\includegraphics[scale=0.22]{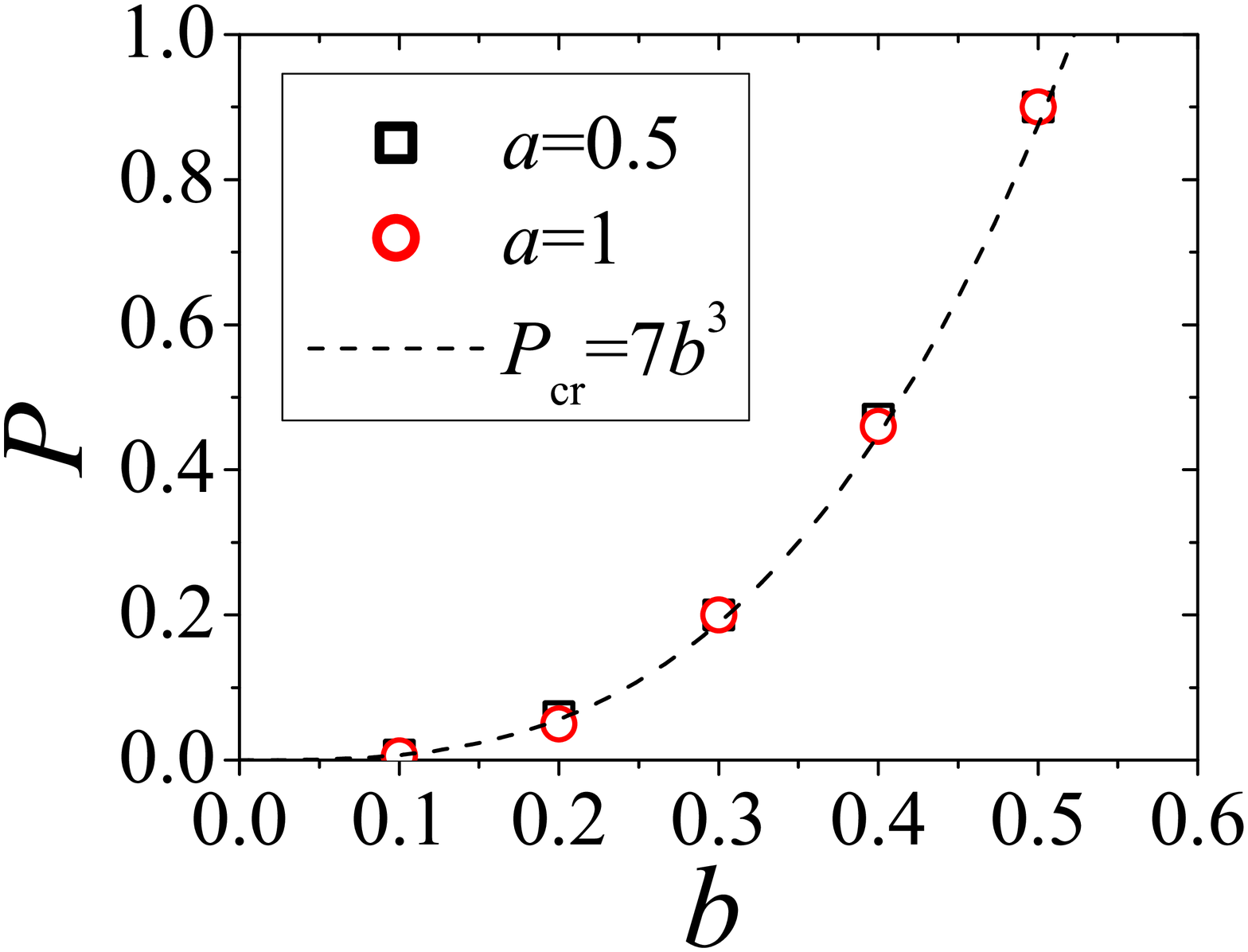}}
\caption{(Color online) Examples of stable symmetric (a), asymmetric (b),
and antisymmetric (c) solitons found in the weekly-coupled system [see Fig.
\protect\ref{Fig1}(a)] for $a=1,$ $b=0.4$, and total norm $P=0.4$ (a) and $%
P=0.6$ (b,c) The antisymmetric solitons are stable below the boundary, $%
P\approx 7b^{3}$, shown in the plane of $\left( b,P\right) $ in panel (d),
at different values of $a$.}
\label{SyandAsy}
\end{figure}

Typical examples of stable symmetric, asymmetric, and antisymmetric solitons
found in the system represented by Fig. \ref{fig1a} are displayed in Fig. %
\ref{SyandAsy}. Naturally, the asymmetric solitons, morphed by the stronger
nonlinearity, are narrower and taller than their symmetric counterparts.

It is relevant to mention that, unlike the standard model of the coupler
with the local cubic nonlinearity, where symmetric and antisymmetric
solitons are available in an obvious exact form, and asymmetric ones can be
effectively described by means of the variational approximation \cite%
{fibers2,Progress}, in the present strongly nonlocal system such an
approximation is not practically possible. Nevertheless, some results can be
obtained in an approximate analytical form for the present system too, see
Eqs. (\ref{scaling}) and (\ref{scaling2}) below.

Before proceeding to the consideration of the transition between the
symmetric and asymmetric solitons, we display the stability area of the
antisymmetric ones (which do not undergo any \textit{antisymmetry-breaking}
bifurcation), in the plane of $\left( b,P\right) $, in Fig. \ref{fig2c}. The
stability boundary may be fitted to curve
\begin{equation}
P=7b^{3}.  \label{cubic}
\end{equation}
Unstable antisymmetric solitons gradually decay into radiation (not shown
here in detail).

The SSB for solitons is summarized in Fig. \ref{SyandAsy} by means of the
bifurcation diagrams, which clearly show that the symmetry-breaking
bifurcation, driven by the nonlocal attractive DDIs, is supercritical, on
the contrary to the commonly known subcritical bifurcation in the coupler
with the local self-focusing nonlinearity \cite{fibers,fibers2}. A similar
trend to the change of the character of the SSB bifurcation for solitons
from sub- to supercritical, with the increase of the degree of the
nonlocality, was recently demonstrated in the model of the optical coupler
with the weakly nonlocal thermal nonlinearity \cite{Fangwei}.

\begin{figure}[tbp]
\centering\subfigure[] {\label{fig3a}
\includegraphics[scale=0.2]{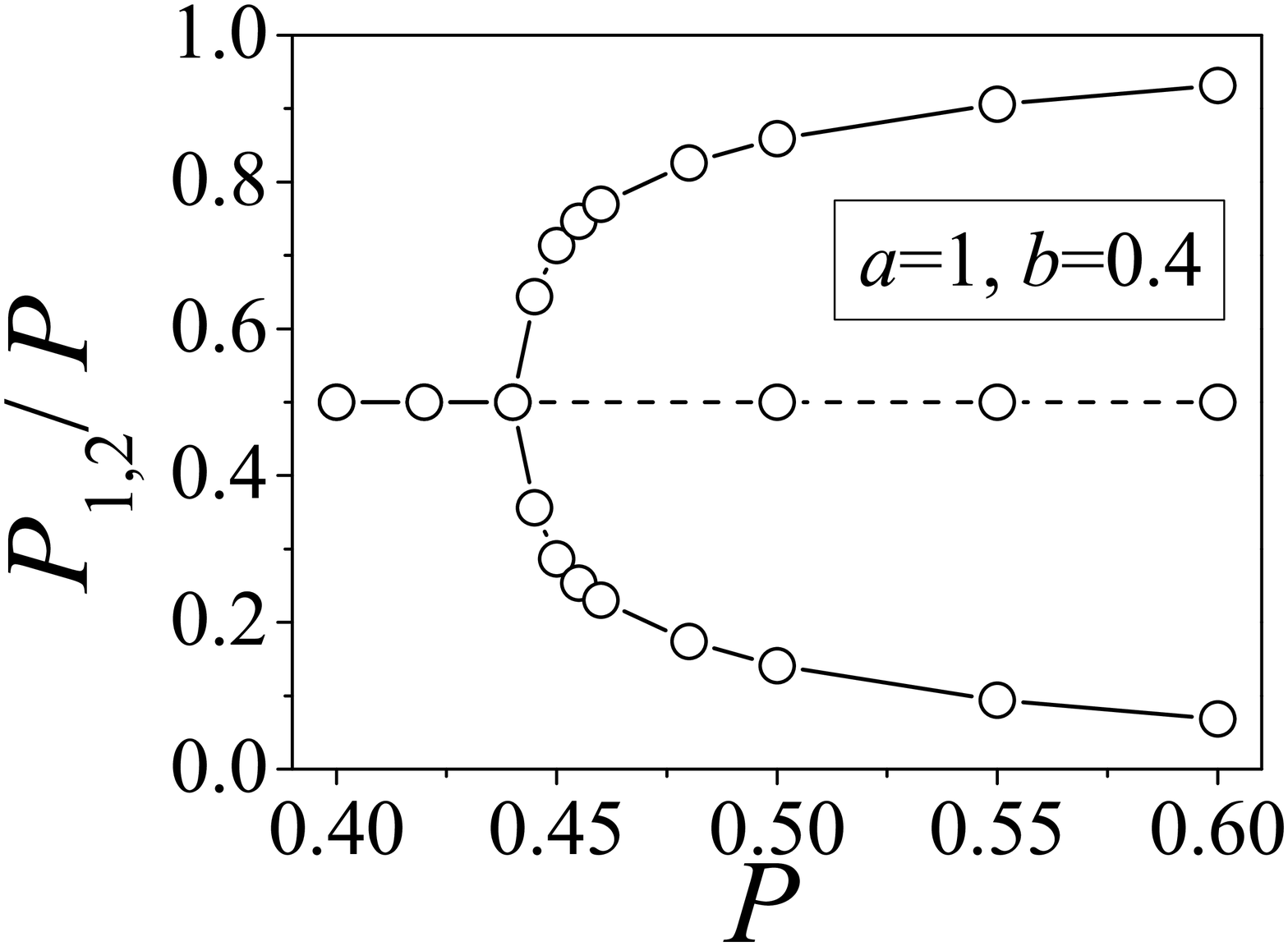}}%
\subfigure[] {\label{fig3b}
\includegraphics[scale=0.2]{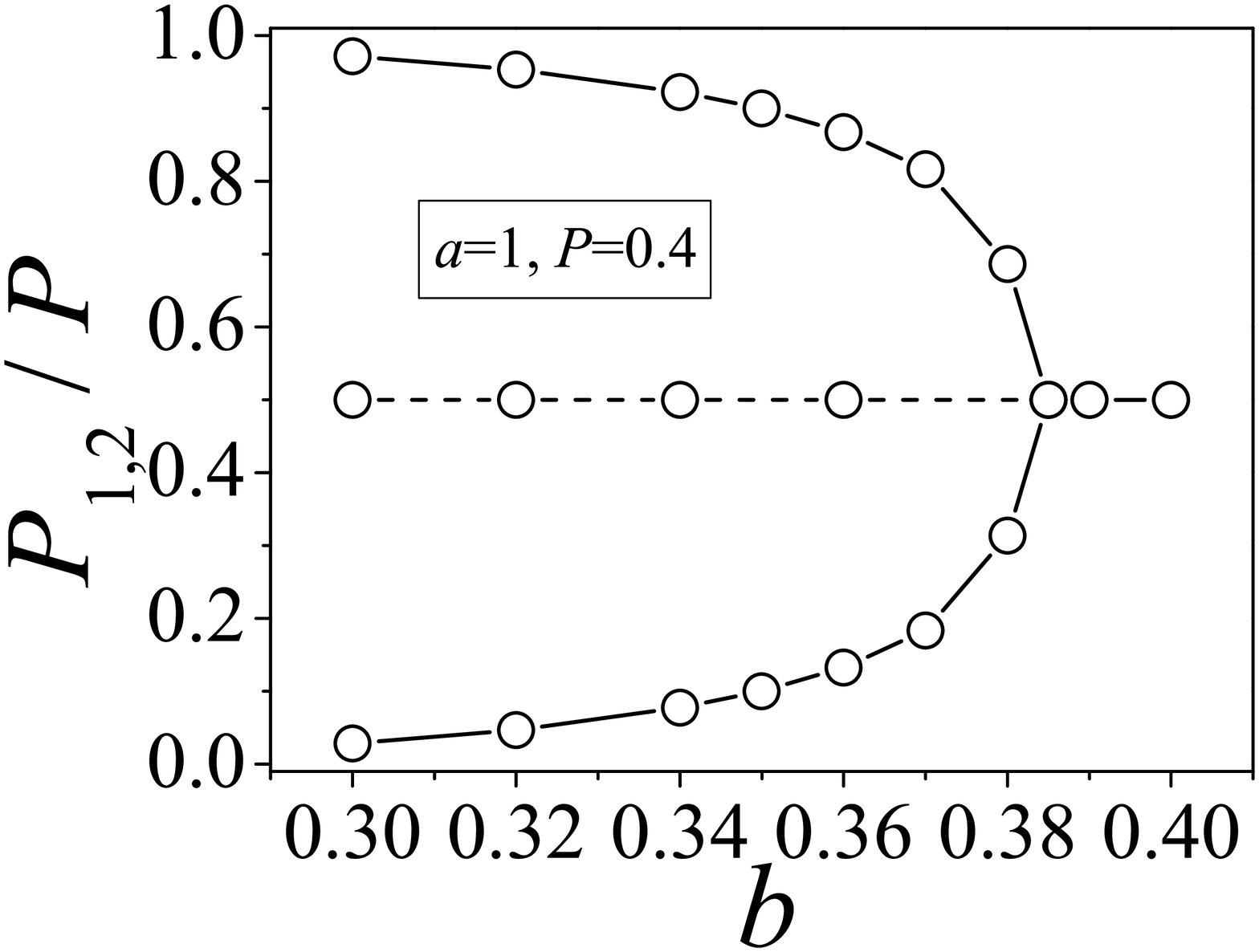}}
\subfigure[] {\label{fig3c}
\includegraphics[scale=0.2]{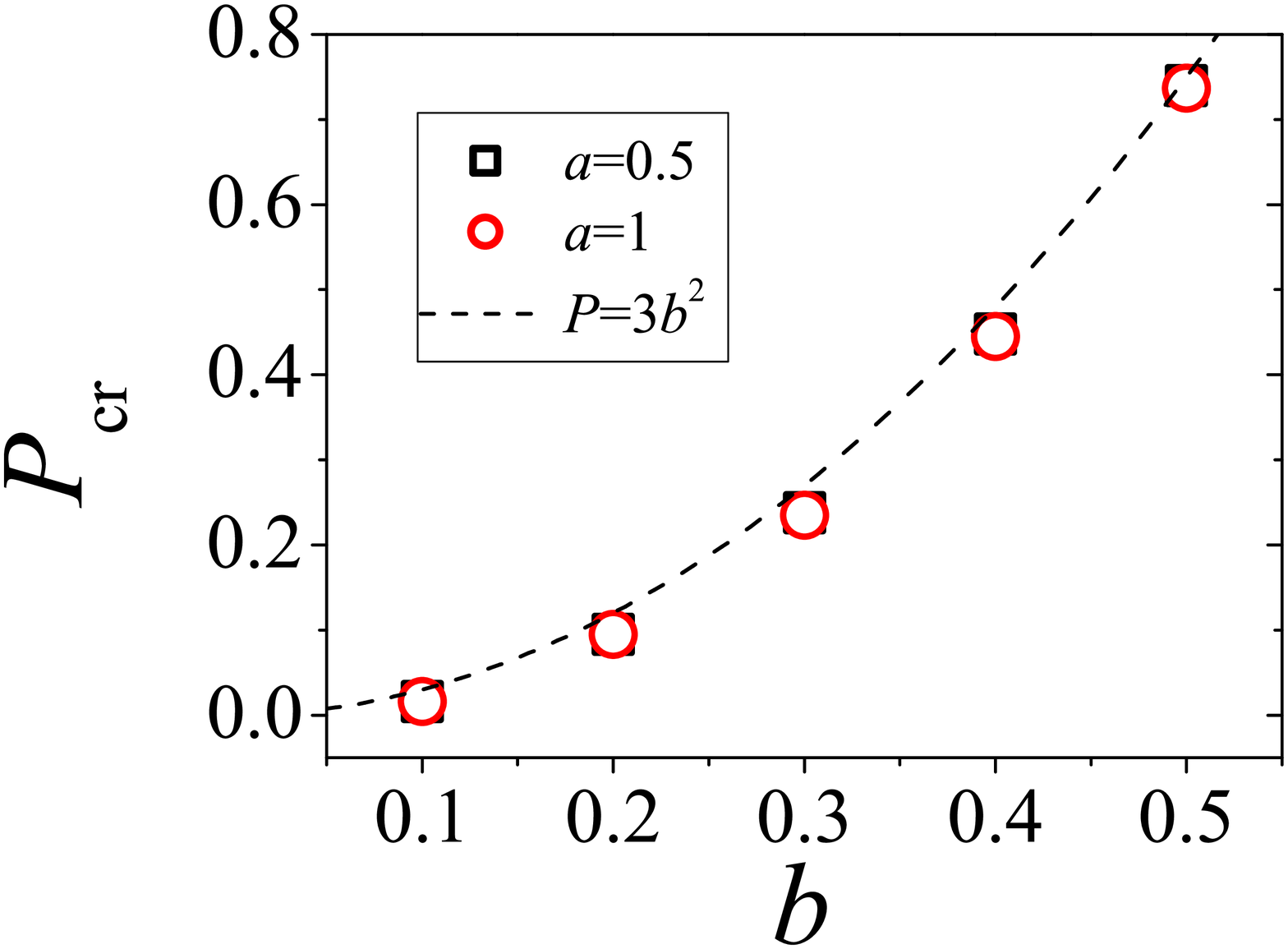}}
\caption{(Color online) (a) The bifurcation diagram for solitons in the
weakly-coupled system: The soliton's asymmetry, measured by the deviation of
the share of the total power in one core ($P_{1}/P$) from $0.5$, versus
total norm $P$. (b) The bifurcation diagram as a function of the pipes'
diameter, $b$. The circles located along the solid and dashed lines
represent stable and unstable solutions, respectively. (c) The critical
value of the total norm at the symmetry-breaking point as a function of $b$,
for fixed $a=1$. Stable asymmetric solitons exist above the curve, while the
symmetric solitons are stable below it. The curve is well fitted by $P_{%
\mathrm{cr}}=3b^{2}$.}
\label{Bifurcation}
\end{figure}

As seen from the structure of the first term in the integrand of Eq. (\ref%
{NLS}), the decrease of diameter $b$ of the parallel pipes implies effective
enhancement of the nonlinearity. This, as well as the direct strengthening
of the nonlinearity due to the increase of the total norm ($P$), leads to
the symmetry breaking, as seen in Figs. \ref{fig3b} and \ref{fig3a}.
Further, Fig. \ref{fig3c} demonstrates the related effect of the decrease of
the critical value, $P_{\mathrm{cr}}$, of the total norm at the bifurcation
point with the decrease of $b$. The latter dependence may be fitted to
formula\textbf{\ }$P_{\mathrm{cr}}=3b^{2}$, which is explained by the fact
that, at small $b$, the nearly diverging first integral term in Eq. (\ref%
{phi}) may be estimated as $A^{2}/b$, where $A$ is the soliton's amplitude.
Its balance with other terms in the equation leads to estimates for scalings
of the amplitude and width:
\begin{equation}
A\sim P/b,W\sim b^{2}/P.  \label{AW}
\end{equation}%
Then, as the SSB point is determined by the competition between the
nonlinear intra-core terms and linear inter-core-coupling ones in Eq. (\ref%
{NLS}) \cite{Snyder}-\cite{fibers}, the corresponding scaling for the value
of $P$ at the critical point indeed takes the form of
\begin{equation}
P_{\mathrm{cr}}\sim \sqrt{\kappa }b^{2}  \label{scaling}
\end{equation}%
(here, $\kappa $ is kept for clarity, although it was actually scaled to be $%
\kappa \equiv 1$).

Note that, according to Eq. (\ref{cubic}), the stability region for
antisymmetric solitons is much smaller at small $b$. This agrees with the
general trend of the antisymmetric solitons in nonlinear couplers to be more
fragile modes than their symmetric counterparts \cite{fibers,PT,PT2}, due to
the obvious fact that they correspond to a larger coupling energy. In fact,
the cubic scaling in Eq. (\ref{cubic}) may be qualitatively explained too,
although in a more vague form than Eq. (\ref{scaling}). Indeed, the
antisymmetric soliton with amplitude $A$ is subject to an oscillatory
instability characterized by complex growth rates (eigenvalues), $\lambda
=\pm i\kappa +\mathrm{Re}\left( \lambda \right) $, where, in the generic
case, an estimate $\mathrm{Re}\left( \lambda \right) \sim A$ is valid (see,
e.g. Ref. \cite{PT2}). Because the instability is oscillatory, the
corresponding perturbations tend to escape from the region of width $W$ ,
occupied by the soliton [see Eq. (\ref{AW})], within time $\tau \sim W/V_{%
\mathrm{gr}}$, where the group velocity is determined by the characteristic
wavenumber of the perturbation mode, $k\sim W^{-1}$, i.e., $\tau \sim W^{2}$%
. The instability accounted for by the escaping perturbations seems as
convective one, that will have enough time to destroy the antisymmetric
soliton under condition $\mathrm{Re}\left( \lambda \right) \tau \sim 1$,
i.e., according to the above estimates,%
\begin{equation}
AW^{2}\sim b^{3}/P\sim 1,  \label{scaling2}
\end{equation}%
which qualitatively explains the numerically found fit (\ref{cubic}).

It is also worthy to note that additional analysis demonstrates that, as is
strongly suggested by Figs. \ref{fig2d} and \ref{fig3c}, the general picture
of the SSB, being sensitive to the value of $b$, shows little dependence on
the separation between the cores, $a$ (in the case of $a>b$, which is
considered here). In other words, the SSB in the weakly-coupled system, in
accordance with its name, is weakly sensitive to the DDI between the
parallel cores, which renders the picture relatively simple.

\subsection{The strongly-coupled system ($a<b$)}

In the setting displayed in Fig. \ref{fig1b}, the small separation between
the effective cores makes effects of the inter-core DDIs essentially
stronger, in comparison with the weakly-coupled system. In fact, the
strongly-coupled system realizes an example of \textit{competing interactions%
}, namely, intra-core attraction and inter-core repulsion. A somewhat
similar example is the discrete Salerno model with competing signs of the
onsite and intersite cubic nonlinearities, which was studied in 1D and 2D
settings \cite{Zaragoza}.

\begin{figure}[tbp]
\centering\subfigure[] {\label{fig4a}
\includegraphics[scale=0.23]{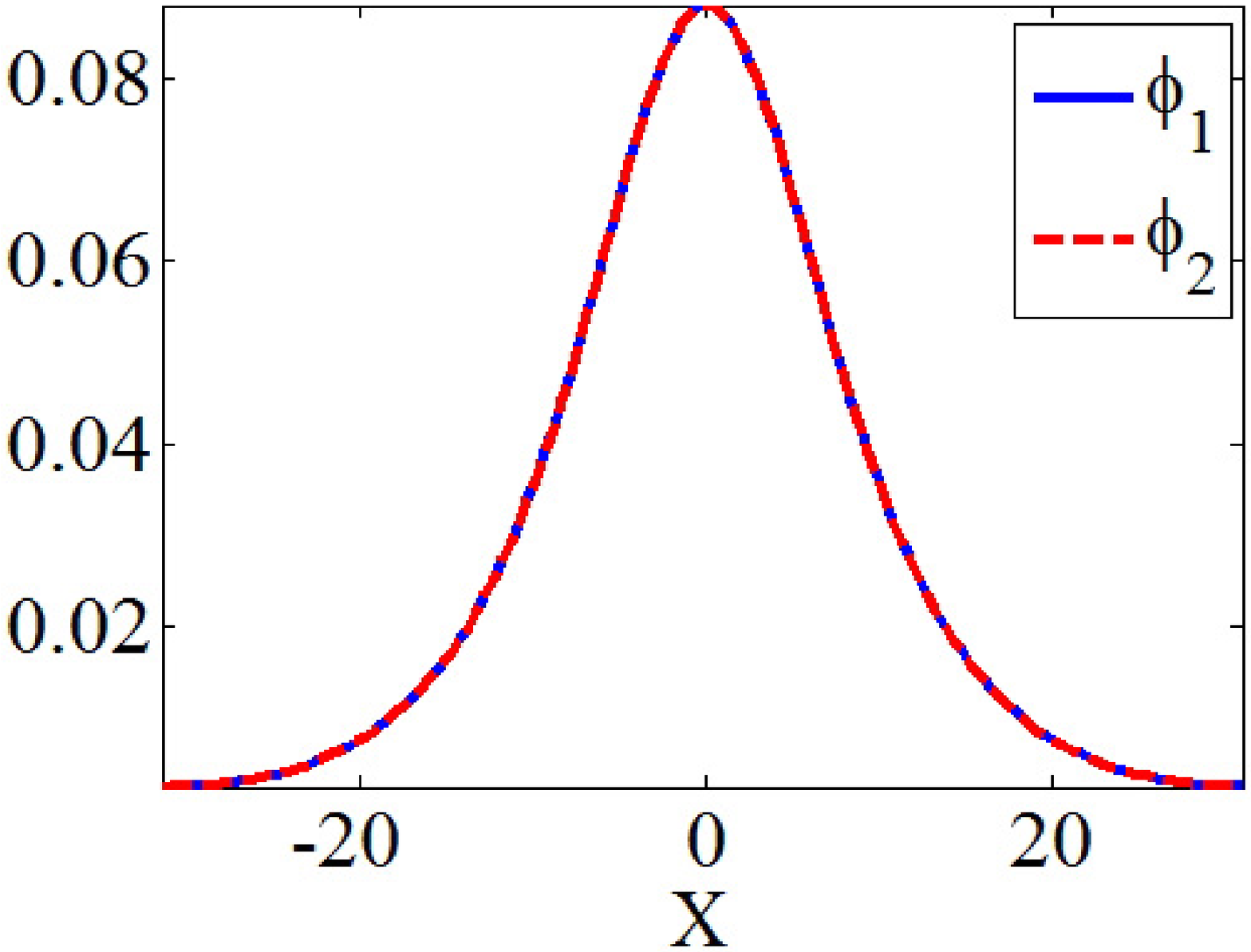}}%
\subfigure[] {\label{fig4b}
\includegraphics[scale=0.23]{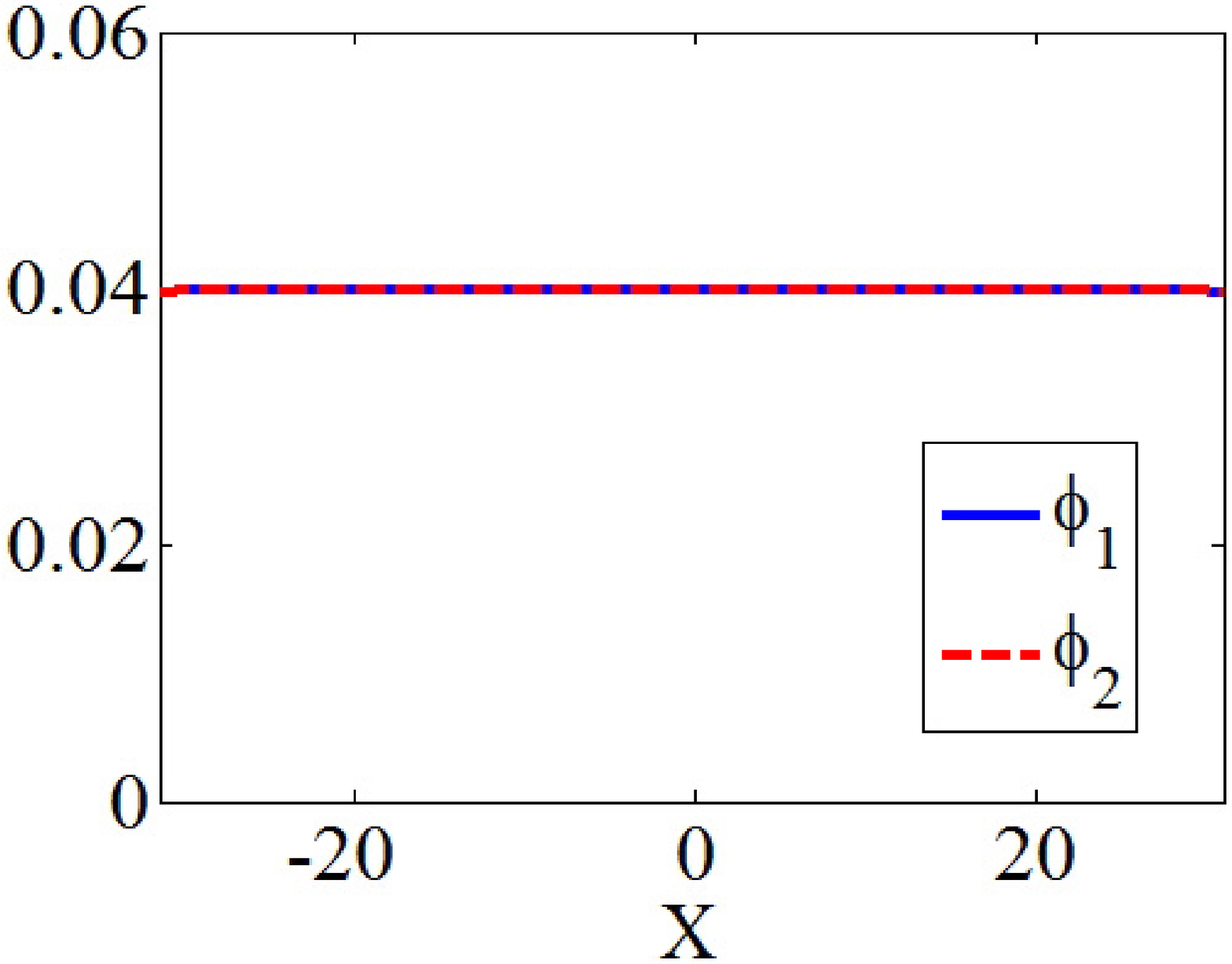}}
\subfigure[] {\label{fig4c}
\includegraphics[scale=0.23]{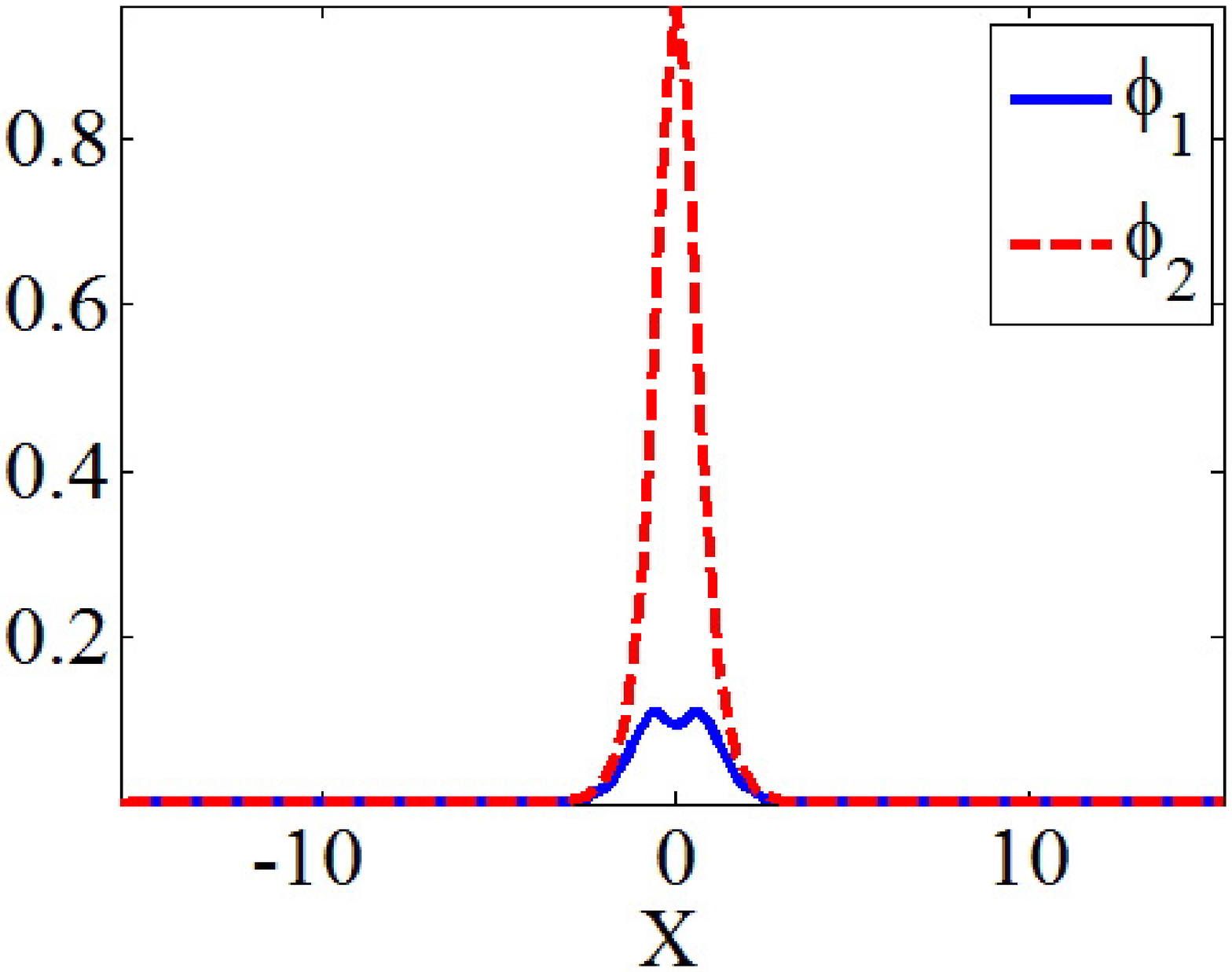}}
\caption{(Color online) Examples of stable modes found in the
strongly-coupled system. (a) A symmetric soliton , with parameters $%
(a,b,P)=(0.3,0.5,0.2)$. (b) A symmetric flat state with $%
(a,b,P)=(0.1,0.5,0.2)$. (c) An asymmetric soliton, with $(a,b,P)=(0.2,0.5,1)$%
. }
\label{smallAstate}
\end{figure}

The numerical solution demonstrates that, in addition to symmetric and
asymmetric solitons, stable \textit{flat states} with unbroken symmetry
between the cores also exist in the strongly-coupled system, being
stabilized by the strong dipolar repulsion between the cores. A typical
example of such stable states is displayed in Fig. \ref{smallAstate}.

In comparison to their counterparts in the weakly coupled system (cf. Fig. %
\ref{SyandAsy}), the symmetric solitons are wider in the case of the strong
coupling. This is also an effect of the repulsive DDI between the cores,
which partly cancels the intra-core attraction, which forms the symmetric
modes. The transition to the asymmetric soliton occurs, as before, with the
increase of the total norm, the difference from the weakly coupled system
being that the smaller component of the asymmetric mode demonstrates a
split-peak structure in Fig. \ref{smallAstate}(b). This feature originates
from the anisotropy [i.e., the $\theta $-dependence in the second term in
the integrand of Eq. (\ref{NLS})] of the DDI between the cores.

\begin{figure}[tbp]
\centering\subfigure[] {\label{fig5a}
\includegraphics[scale=0.18]{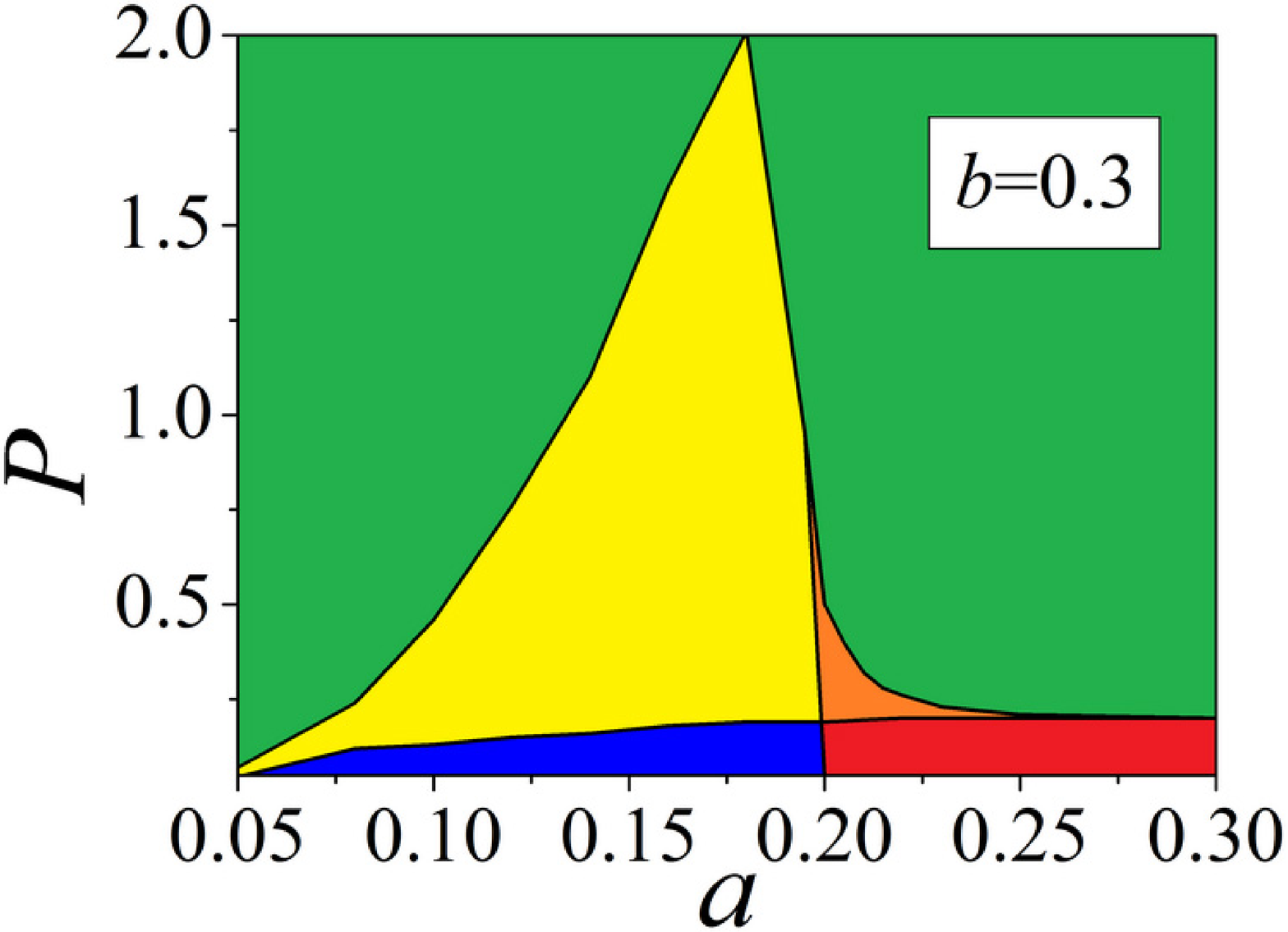}}%
\subfigure[] {\label{fig5b}
\includegraphics[scale=0.18]{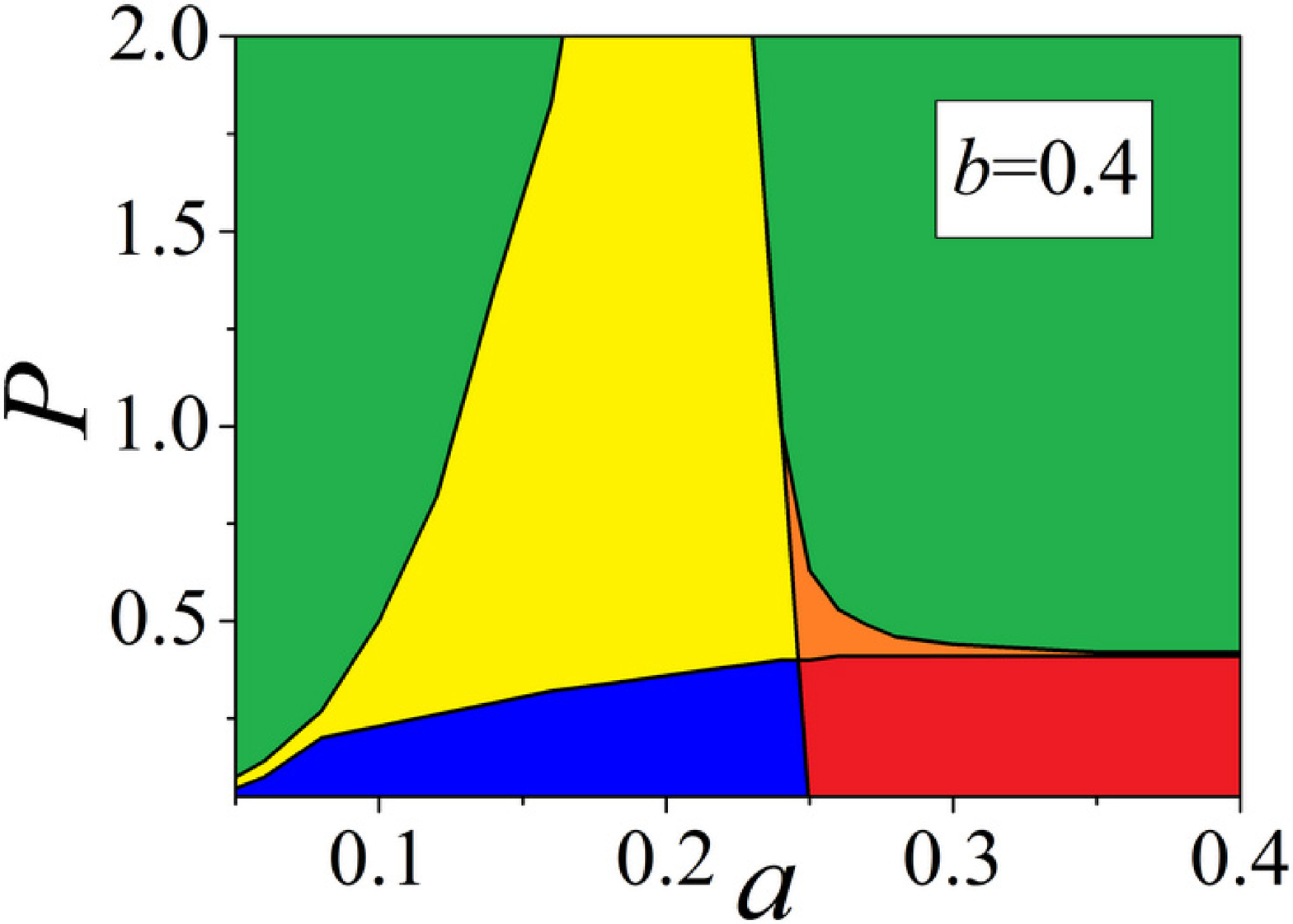}}
\subfigure[] {\label{fig5c}
\includegraphics[scale=0.18]{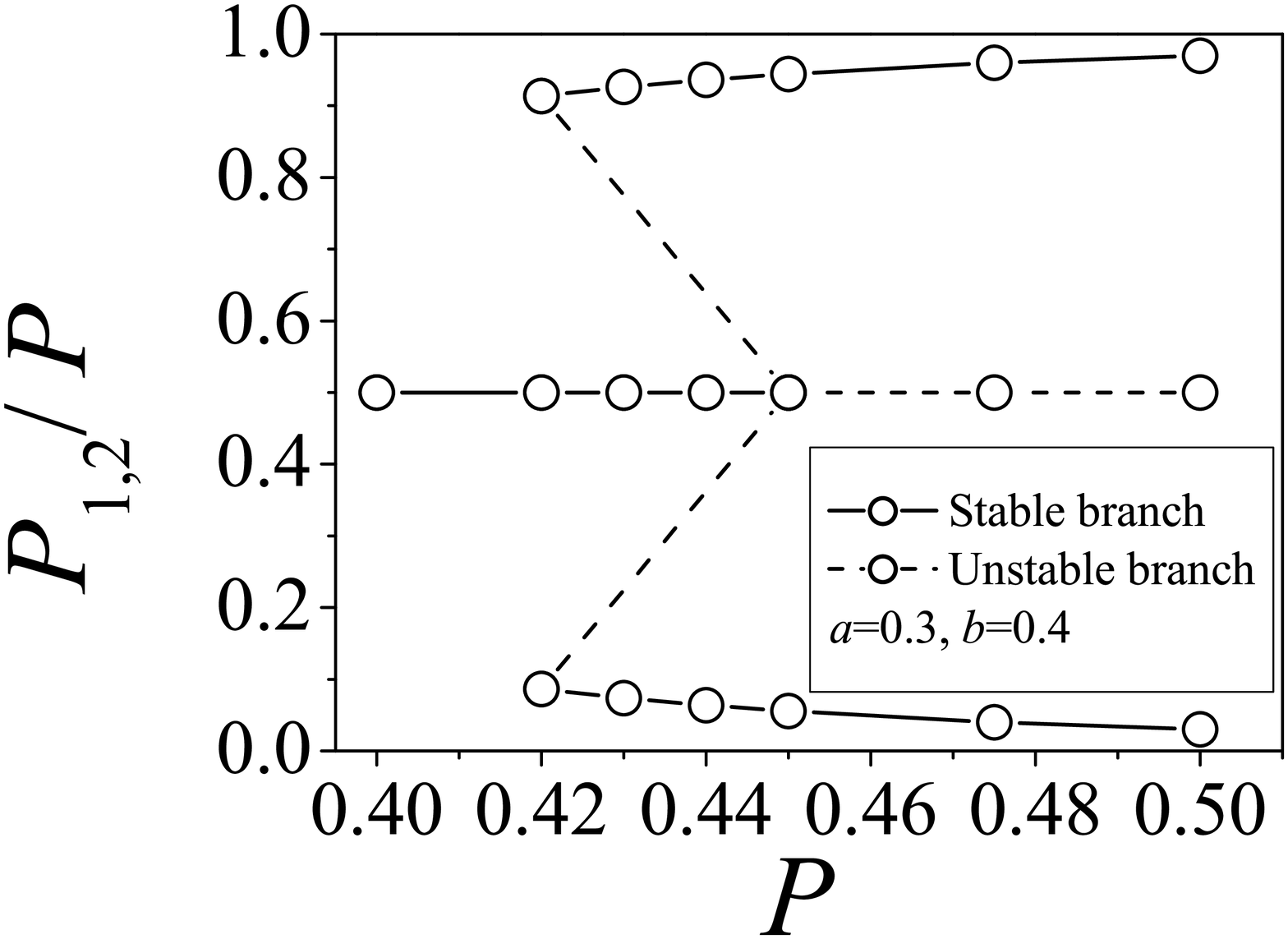}}
\caption{(Color online) (a), (b) Stability diagrams in the plane of the
total norm ($P)$ and separation $a$ between the cores of the
strongly-coupled system, for two different values of the overall diameter, $%
b $. Symmetric solitons are stable in the bottom right areas colored by red,
flat states in the bottom left areas, shown in blue, and asymmetric
solitons---in the green top areas on the left and right sides. The
middle-left yellow areas are regions of the bistability area of flat state
and asymmetric solitons, while the small orange areas at the right center
harbor the bistability of the symmetric and asymmetric solitons. (c) The
bifurcation diagram for solitons in the strongly-coupled system: The
soliton's asymmetry, measured by the deviation of the share of the total
power in one core ($P_{1}/P$) from $0.5$, versus total norm $P$. The type of
the symmetry-breaking bifurcation is subcritical here.The dashed segments,
which link the stable asymmetric branches and the stable symmetric one,
designate the actually missing unstable branches, which, as usual
\protect\cite{Marek}, could not be found by means of the
imaginary-time-integration method.}
\label{Stab1}
\end{figure}

Results of the systematic analysis of the strongly-coupled system are
summarized in Fig. \ref{fig5a} and \ref{fig5b}, in the form of stability
diagrams for all the three types of the modes, in the plane of $\left(
a,P\right) $, for two different fixed values of $b$. There are two
bistability areas, where the stable flat state and asymmetric solitons
coexist [the large yellow triangular and trapezoidal regions in the left
parts of Figs. \ref{fig5a} and \ref{fig5b}, respectively], or the symmetric
and asymmetric solitons are simultaneously stable (small orange regions in
the right parts if the same figures). The presence of the latter bistability
area implies that the SSB transition between the symmetric and asymmetric
solitons is subcritical in the strongly-coupled system, as explicitly shown
in the bifurcation diagram in Fig. \ref{fig5c} [cf. Fig. \ref{fig3a}]. It is
thus concluded that the repulsive DDI between the cores plays an
increasingly more important role with the decrease of the separation between
them, $a,$ which becomes the dominant parameter of the system at $a\lesssim
(2/3)b$. In particular, only in this region the symmetric flat state may be
stable. On the other hand, the effect of parameter $a$ is inconspicuous at $%
a>(2/3)b$, like in the weakly-coupled system, see the previous subsection.
Accordingly, at $a\rightarrow b$, the area of the bistability between the
symmetric and asymmetric solitons (the small orange area in Figs. \ref{fig5a}
and \ref{fig5b}) shrinks to zero, which demonstrates that the SSB changes to
the supercritical type, like in the weekly-couple case. Finally,
antisymmetric solitons and flat states can be also found as stationary
solutions of the strongly-coupled system, but both these species of the
modes turn out to be completely unstable.

\subsection{Effects of the local nonlinearity}

\begin{figure}[tbp]
\centering\subfigure[] {\label{fig6a}
\includegraphics[scale=0.18]{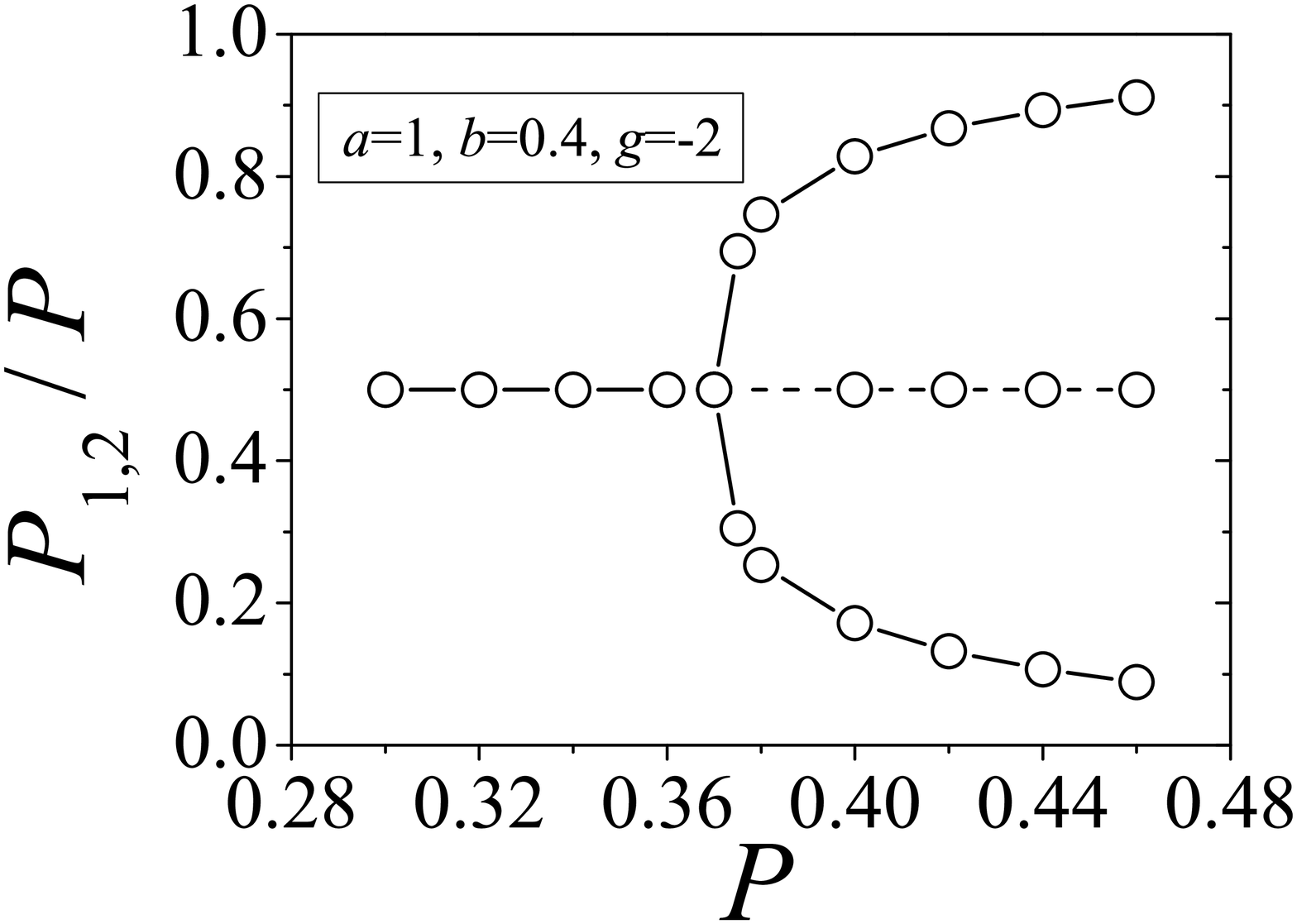}}%
\subfigure[] {\label{fig6b}
\includegraphics[scale=0.18]{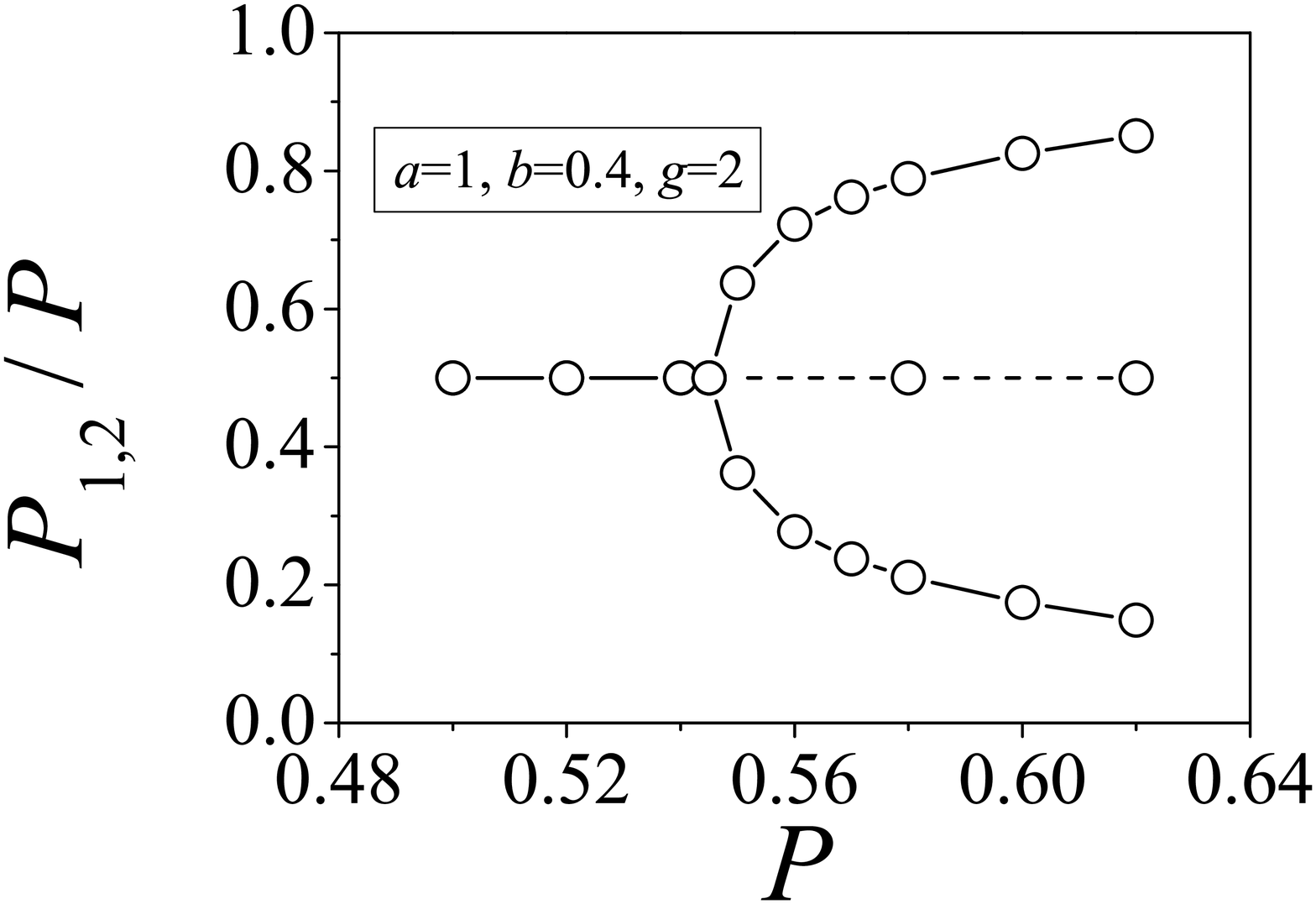}}
\subfigure[] {\label{fig6c}
\includegraphics[scale=0.18]{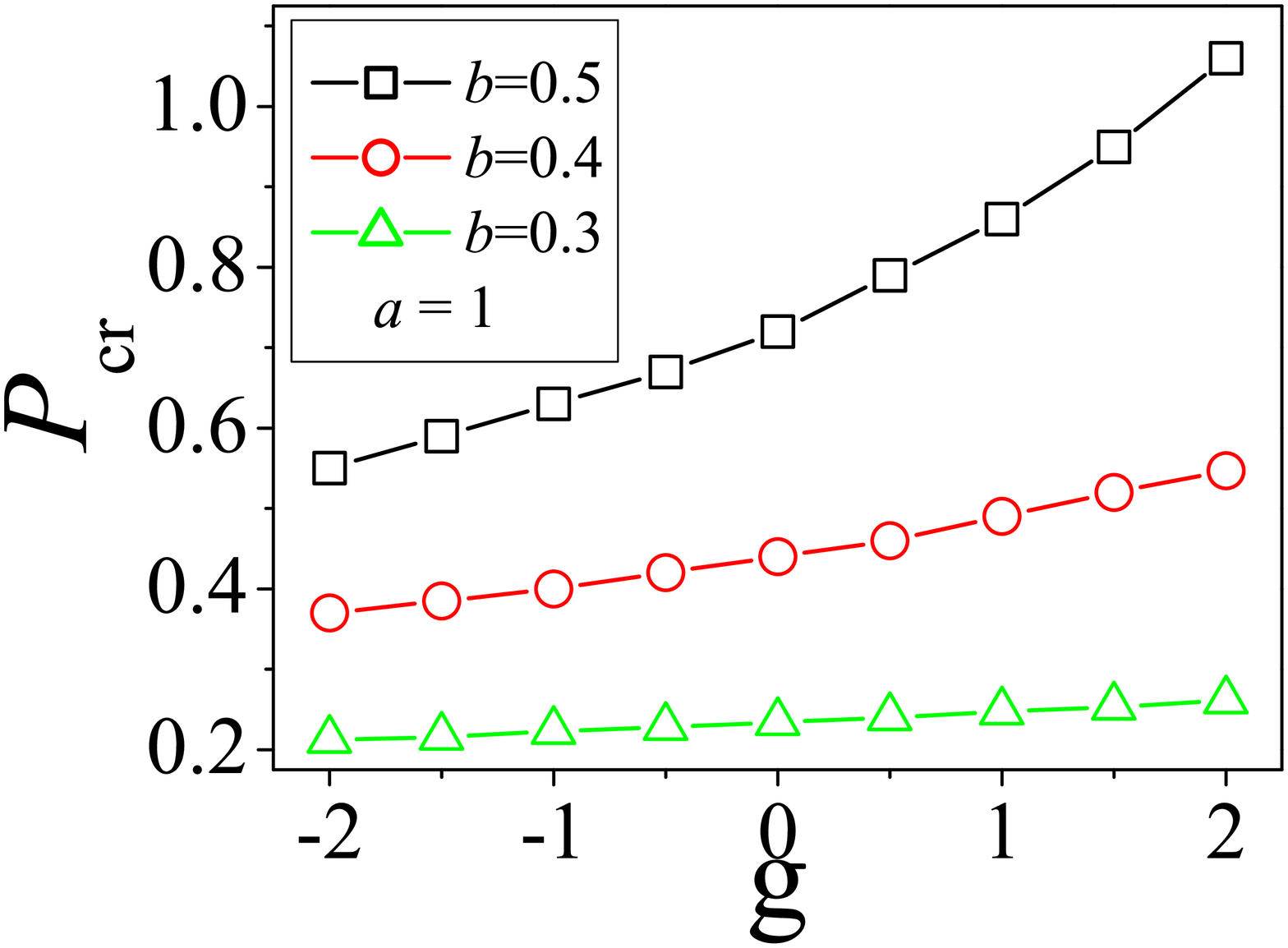}}
\caption{(Color online) The bifurcation diagrams in the weakly-coupled
system, with $(a,b)=(1,0.4)$, in the presence of the self-attractive (a), $%
g=-2$, or self-repulsive (b), $g=+2$, local nonlinearity. (c) The total norm
at the bifurcation point as a function of strength $g$ of the additional
contact nonlinearity, at different fixed values of $b$.}
\label{Stab}
\end{figure}

The inclusion of the local self-attractive ($g<0$) or self-repulsive ($g>0$)
nonlinearity into the coupled GPEs, Eq. (\ref{NLS}), shifts the point of the
SSB bifurcation (the critical value of the total power, $P_{\mathrm{cr}}$),
but it does not change the supercritical character of the bifurcation in the
weakly-coupled system. The bifurcation diagrams for the system of this type,
in the presence of the local nonlinearity of either sign, are presented in
Figs. \ref{Stab}(a) and (b), cf. Fig. \ref{Bifurcation}. In addition, Fig. %
\ref{Stab}(c) displays the dependence of the bifurcation point, $P_{\mathrm{%
cr}}$, on strength $g$ of the local nonlinearity. Naturally, the latter
dependence is weaker for smaller values of the pipes' diameter, $b$, as the
DDI is stronger at smaller $b$, suppressing the effect of the local
nonlinearity. It is natural too that the self-attractive local nonlinearity (%
$g<0$) makes $P_{\mathrm{cr}}$ smaller, while the local self-repulsion ($g>0$%
) makes it larger. Similarly, the addition of the contact nonlinearity does
not change the type of SSB bifurcation in the strongly-coupled system either
(not shown here in detail).

\section{Mobility and collisions between solitons}

\begin{figure}[tbp]
\centering\subfigure[] {\label{fig7a}
\includegraphics[scale=0.25]{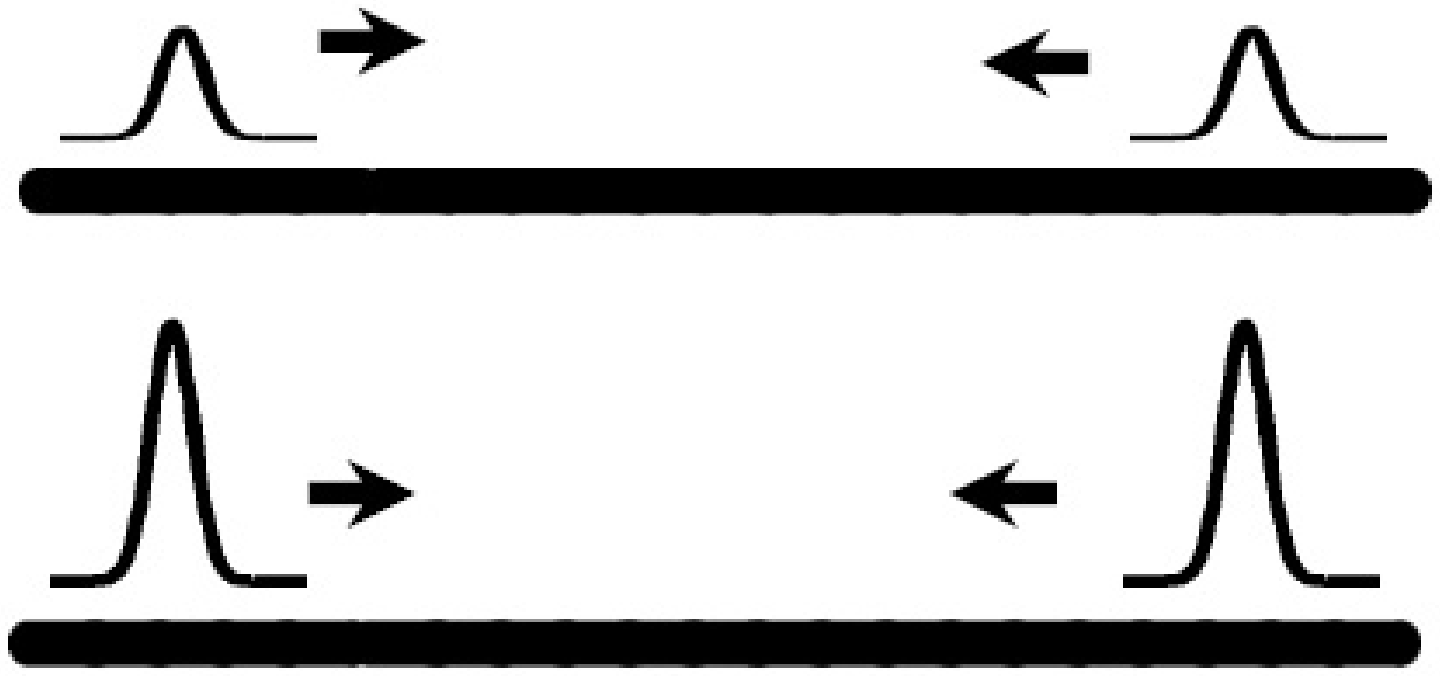}}%
\subfigure[] {\label{fig7b}
\includegraphics[scale=0.25]{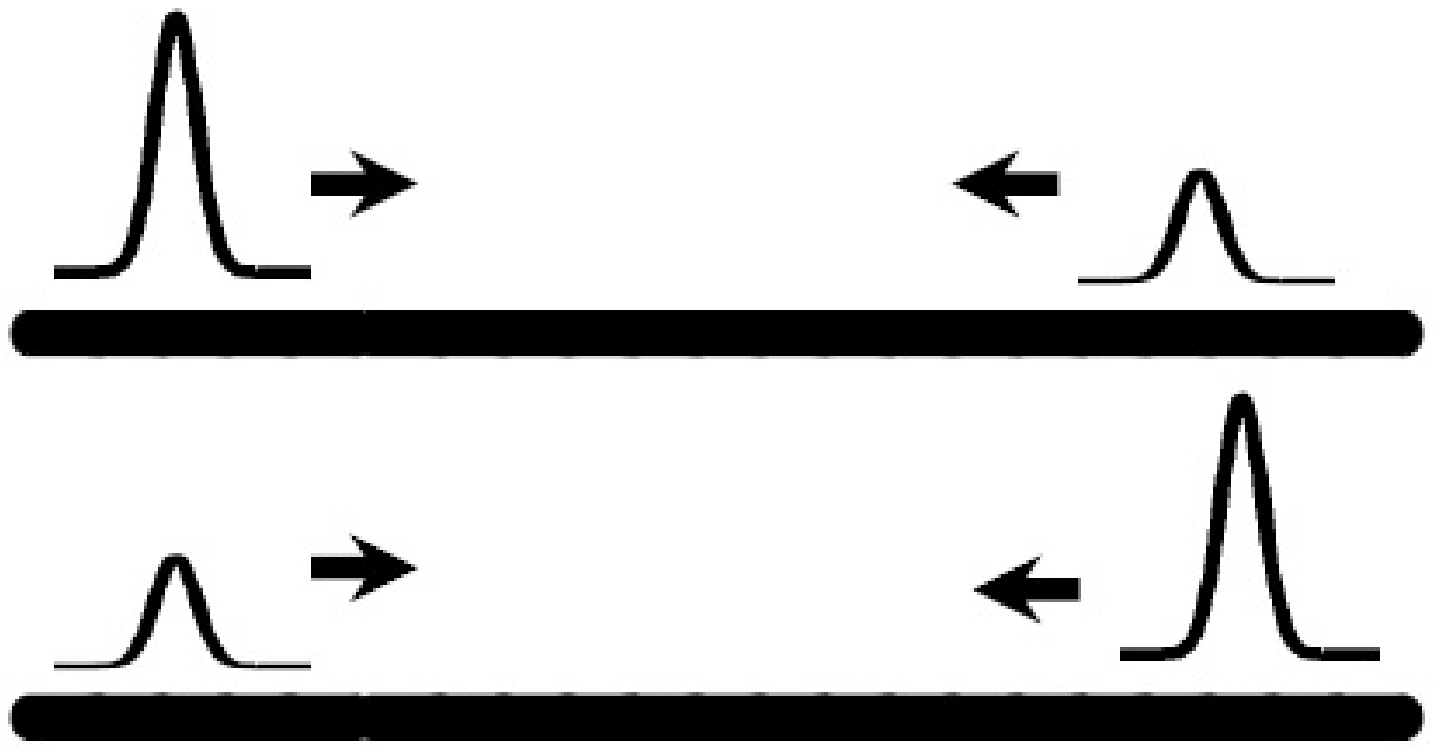}}
\subfigure[] {\label{fig7c}
\includegraphics[scale=0.25]{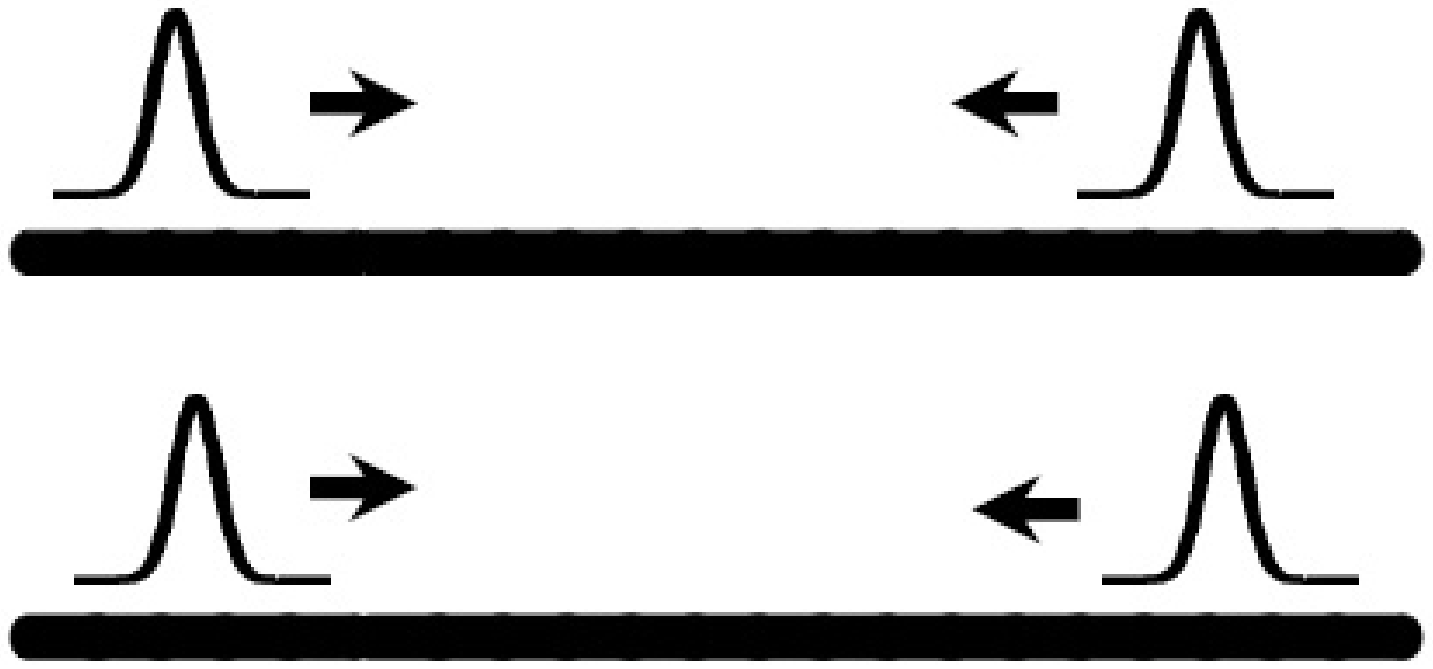}}
\caption{Three types of collisions between solitons in the dual-core system:
(a) ``Unipolar" asymmetric solitons, with the larger components belonging to
the same core. (b) Asymmetric solitons with opposite ``polarities". (c) The
collision between symmetric solitons.}
\label{Colli}
\end{figure}

Collisions between solitons represent an important aspect of the dynamics of
integrable and nonintegrable models \cite{Yang}. In dual-core systems, one
can study collisions between symmetric solitons, and, which is most
interesting, between asymmetric ones with equal or opposite \textit{%
polarities}, i.e., with larger components belonging to the same or different
cores \cite{polarity}. These three cases are schematically defined in Fig. %
\ref{Colli}.

We here report numerical results for collisions between solitons in the
weakly-coupled system, taking the initial state at $t=0$ as a pair of
far-separated kicked solitons,
\begin{equation}
\psi _{1,2}^{(0)}=U_{1,2}^{(1)}(x+x_{0},P)e^{i\eta
(x+x_{0})}+U_{1,2}^{(2)}(x-x_{0},P)e^{-i\eta (x-x_{0})},  \label{0}
\end{equation}%
where $U_{1,2}^{(1,2)}(x\pm x_{0},P)$ are the two-component soliton
solutions with total powers $P$, which are centered at $x=\mp x_{0}$, with
sufficiently large initial separation $2x_{0}$, and $\eta $ is the kick
(momentum imparted to the soliton). Solutions $U^{(1)}$ and $U^{(2)}$ are
either identical ones, or, in the above-mentioned case of opposite
polarities, these are two asymmetric solitons with swapped components, see
Fig. \ref{Colli}(b).

\begin{figure}[tbp]
\centering\subfigure[] {\label{fig8a}
\includegraphics[scale=0.28]{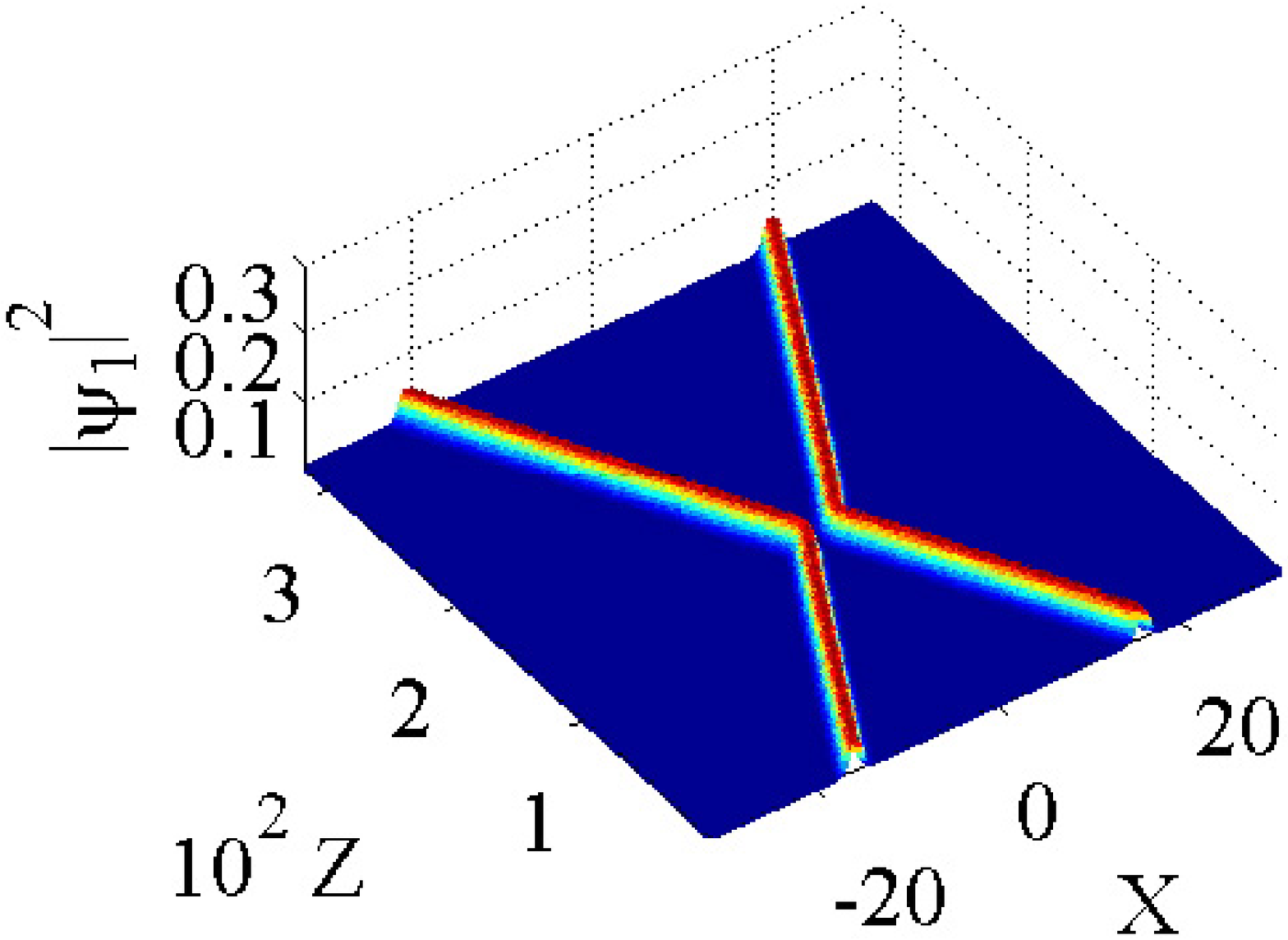}}%
\subfigure[] {\label{fig8b}
\includegraphics[scale=0.28]{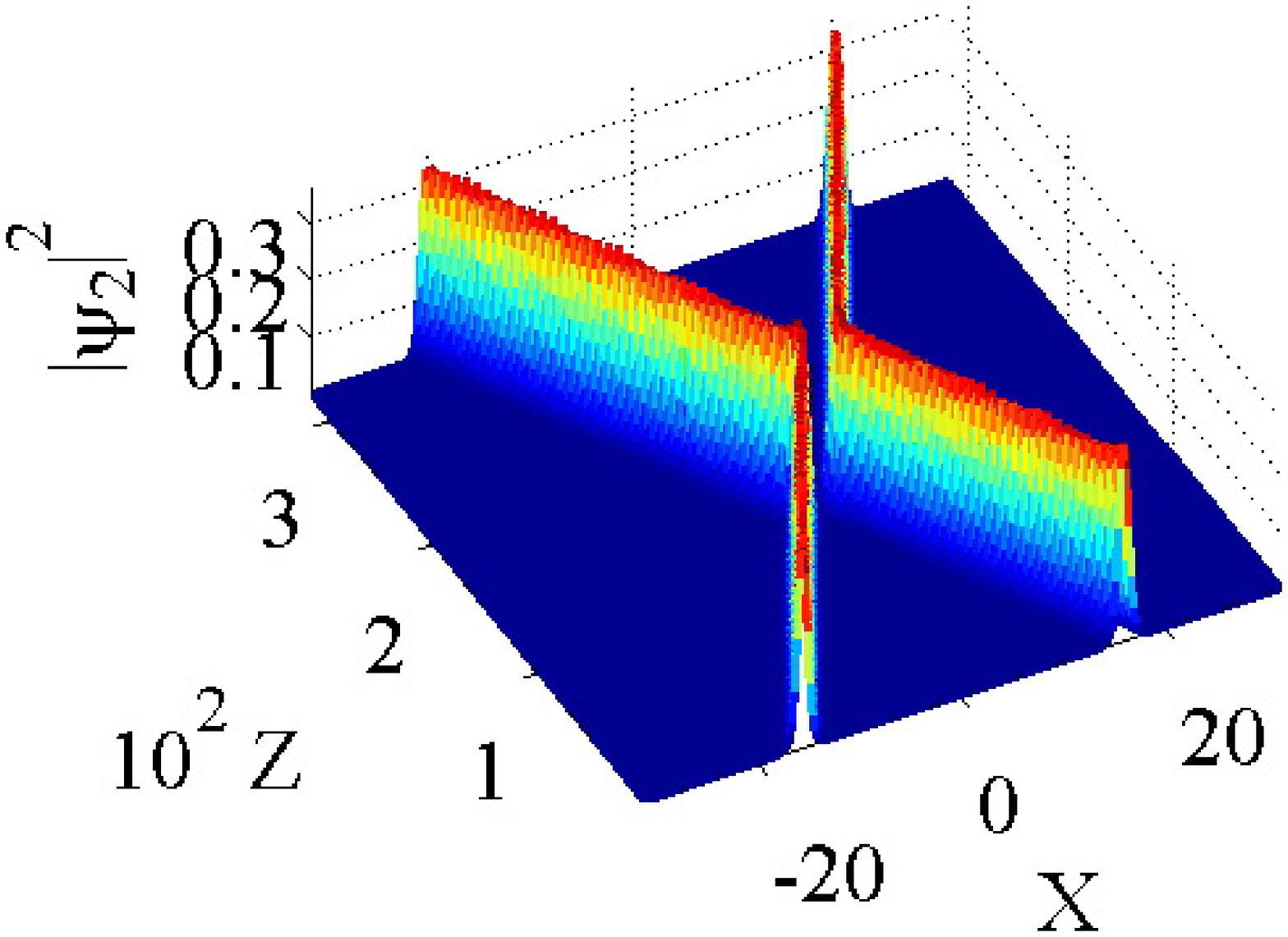}}
\subfigure[] {\label{fig8c}
\includegraphics[scale=0.28]{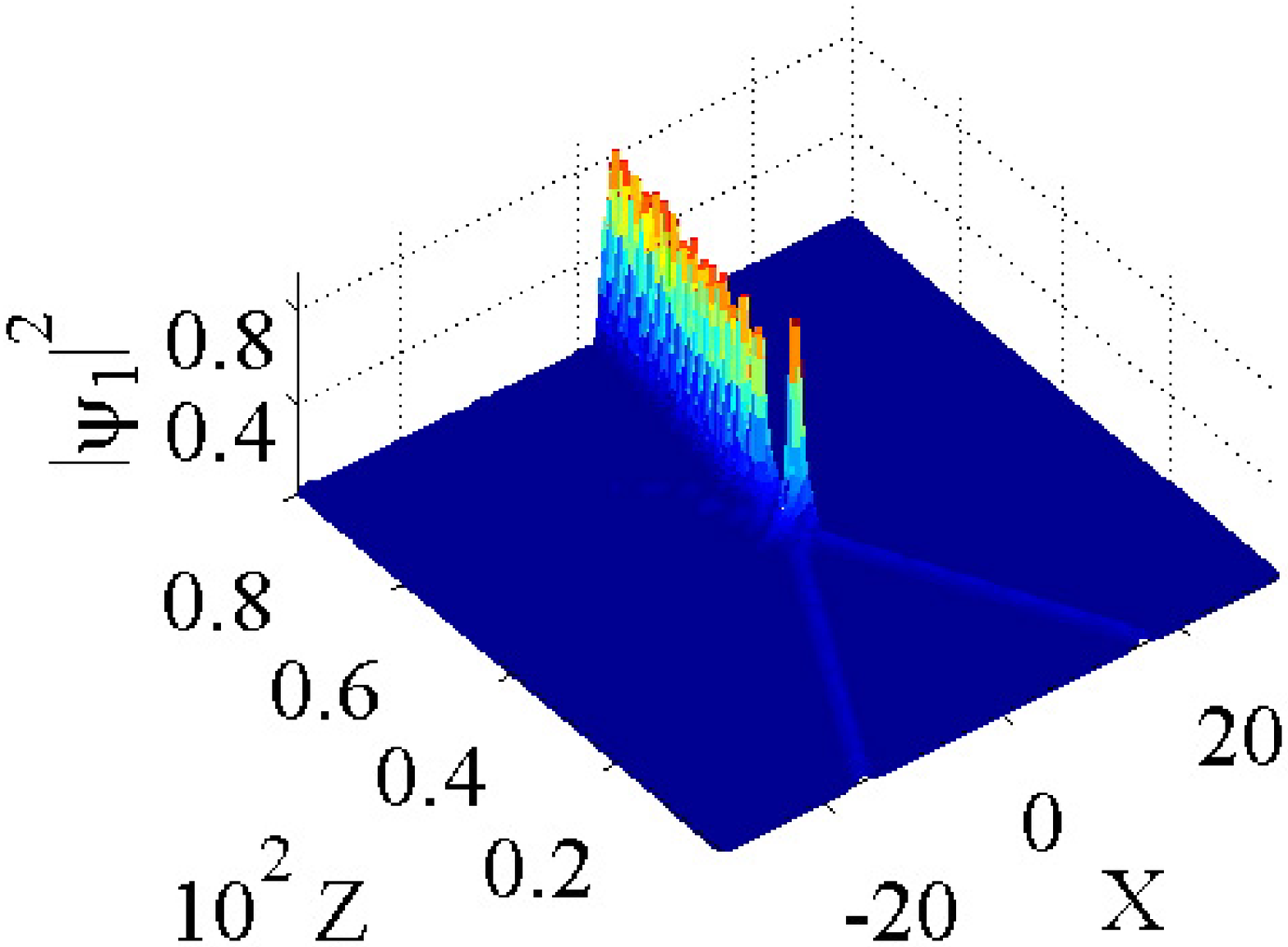}}
\subfigure[] {\label{fig8d}
\includegraphics[scale=0.28]{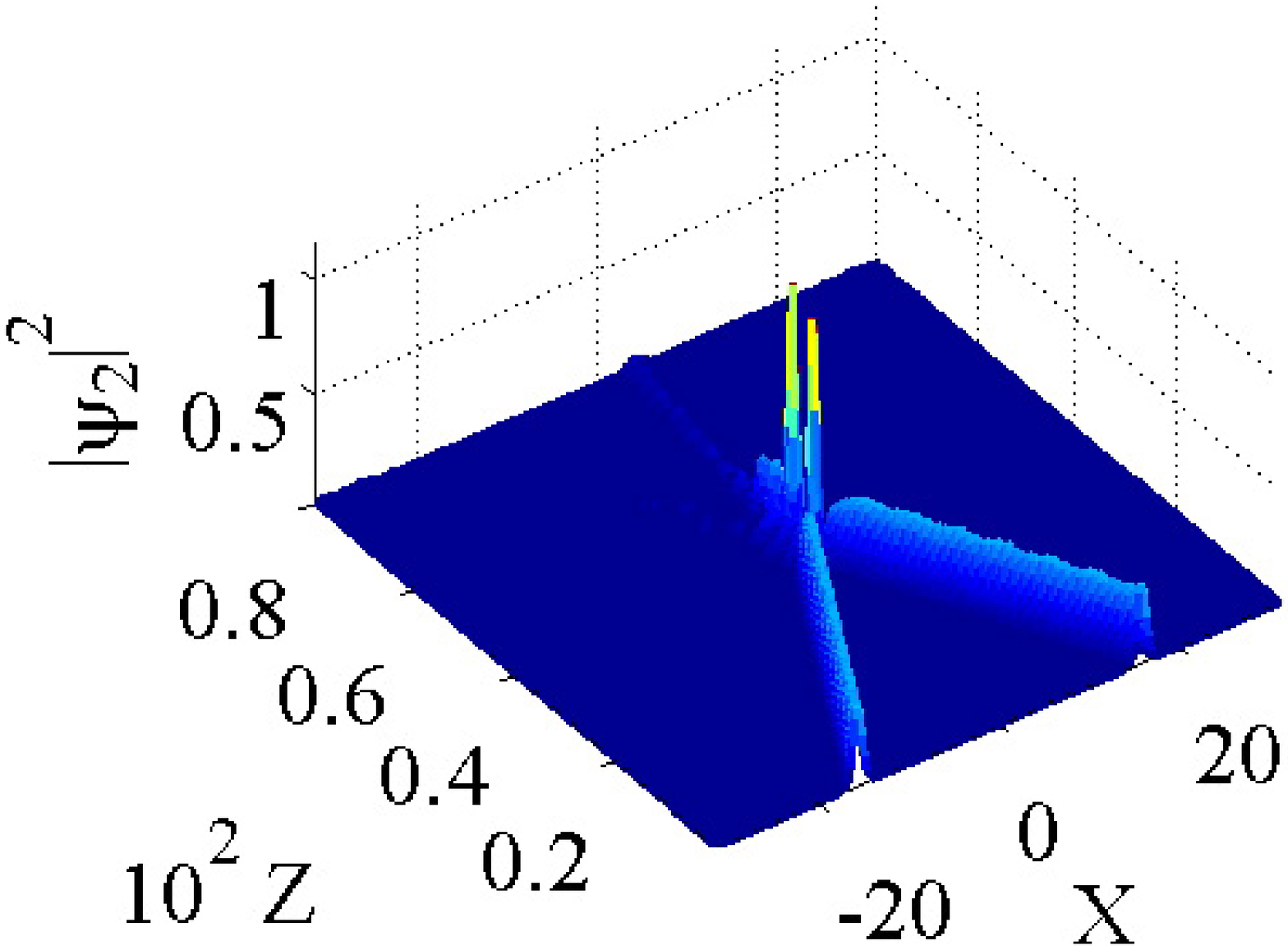}}%
\subfigure[] {\label{fig8e}
\includegraphics[scale=0.28]{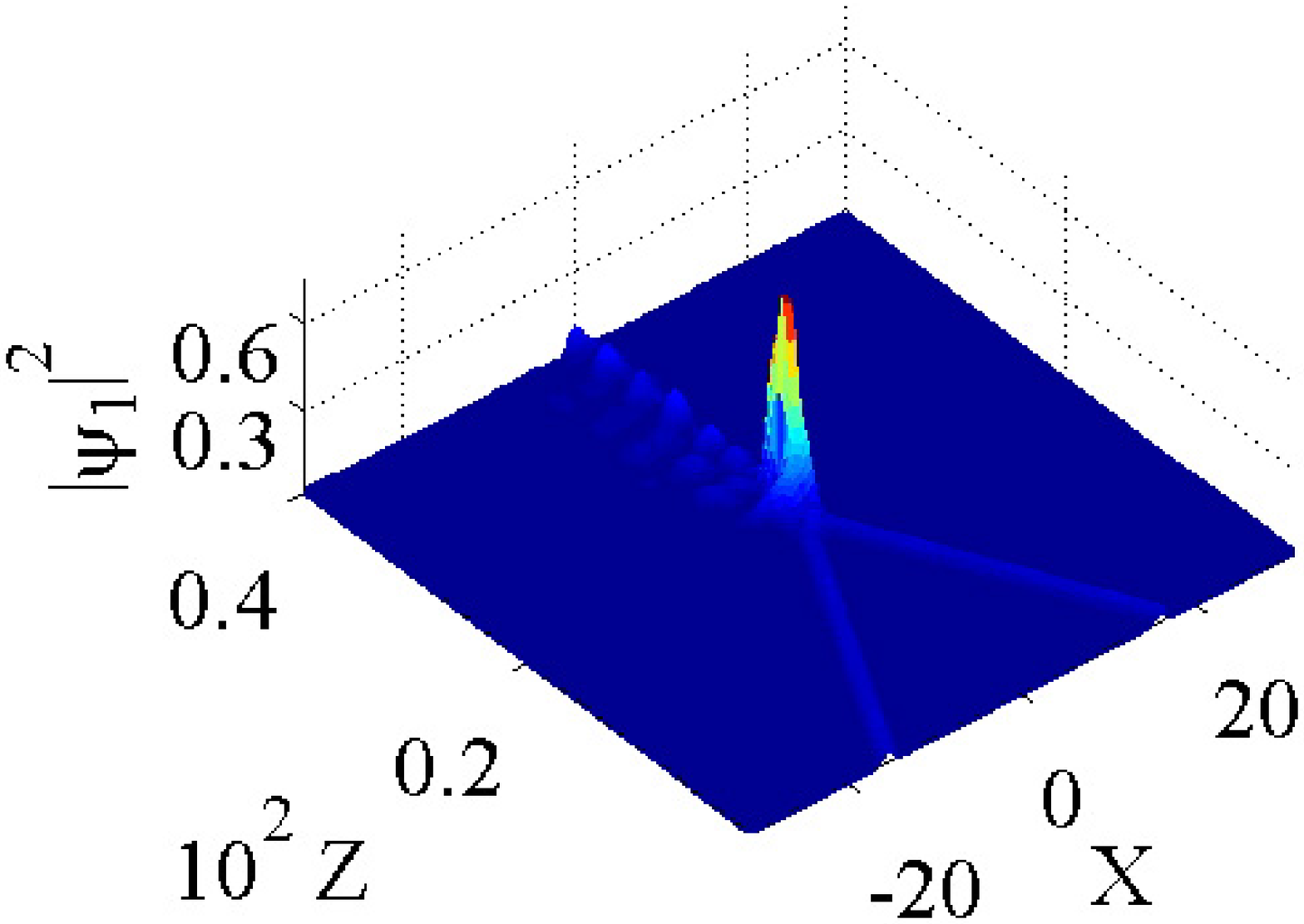}}
\subfigure[] {\label{fig8f}
\includegraphics[scale=0.28]{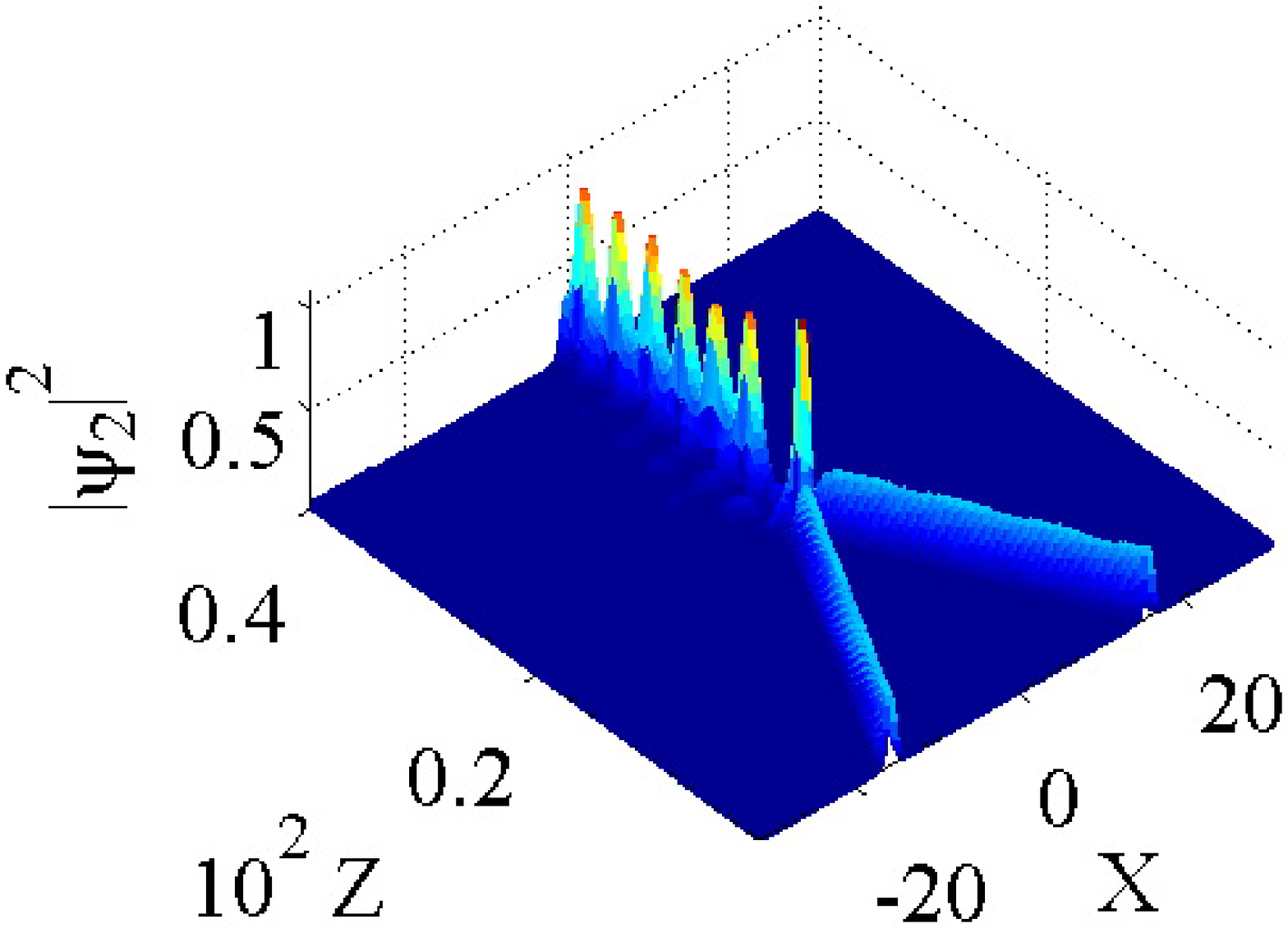}}
\subfigure[] {\label{fig8g}
\includegraphics[scale=0.28]{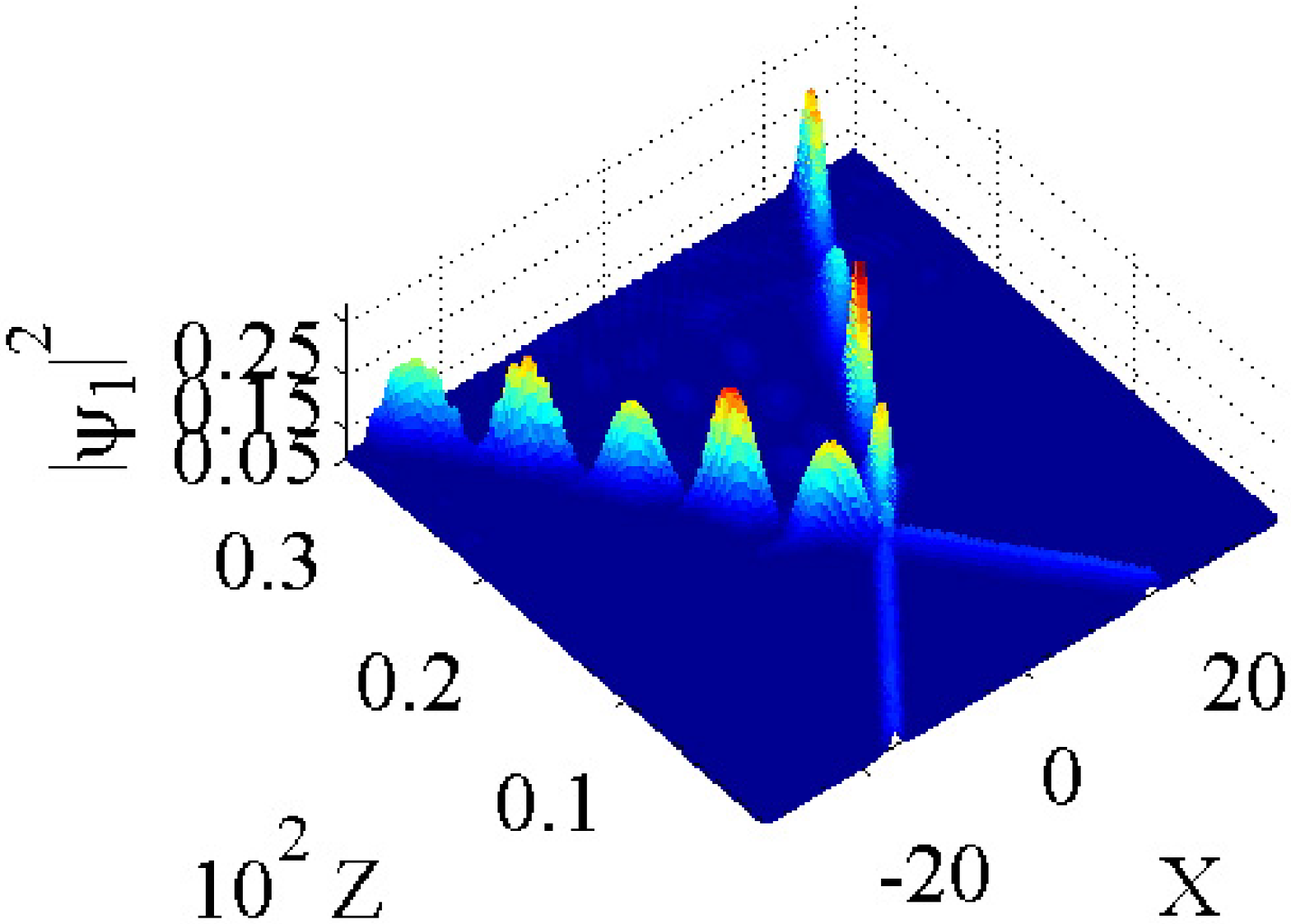}}
\subfigure[] {\label{fig8h}
\includegraphics[scale=0.3]{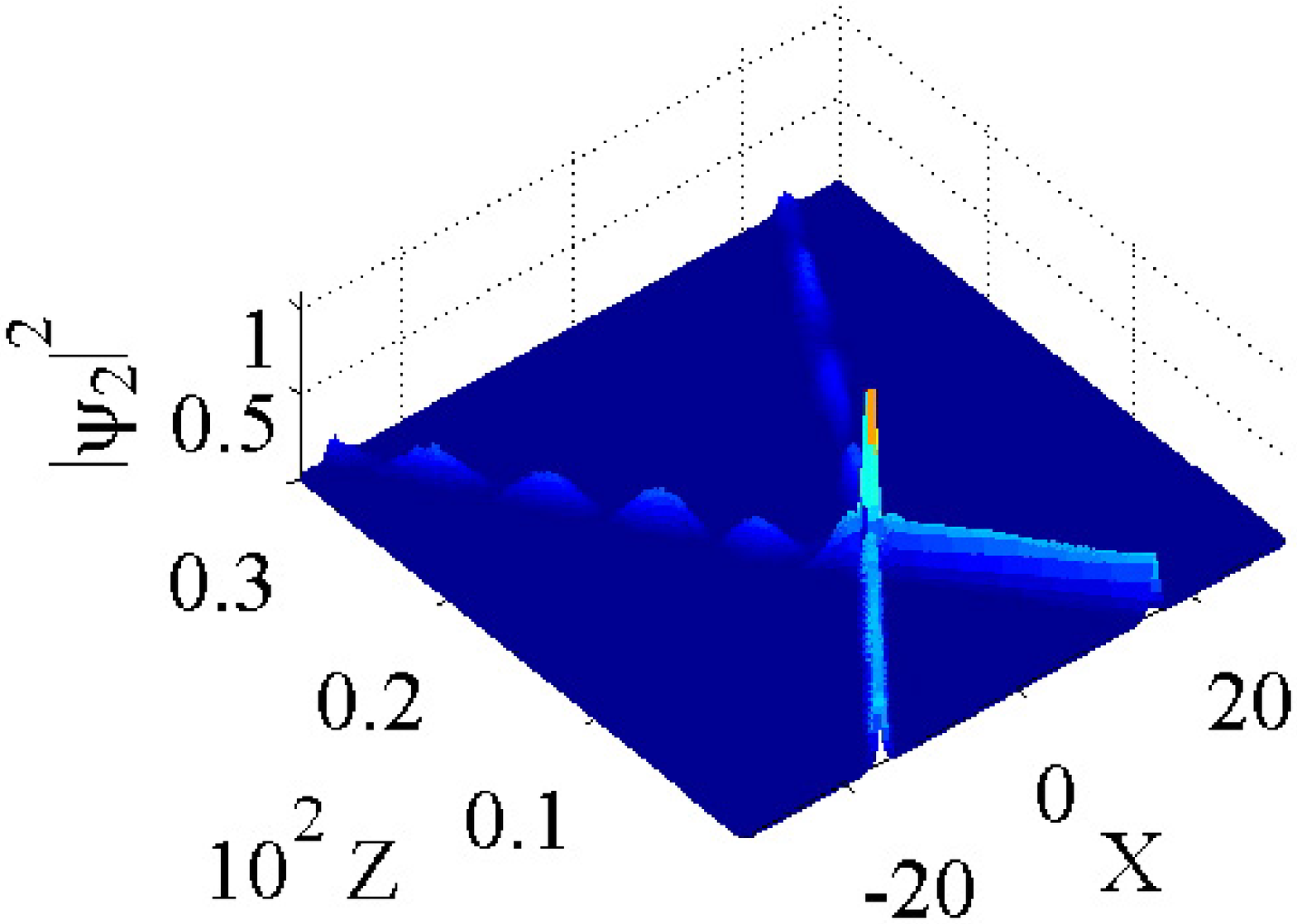}}
\caption{(Color online) Examples of collisions between unipolar asymmetric
solitons: (a,b) for the slow solitons, with $\protect\eta =0.1$; (c,d) for
the intermediate velocity, $\protect\eta =0.4$; (e,f) for a larger
intermediate velocity, $\protect\eta =0.8$; (g,h) for the fast solitons, $%
\protect\eta =1.6$.}
\label{Colli1}
\end{figure}

\begin{figure}[tbp]
\centering\subfigure[] {\label{fig9a}
\includegraphics[scale=0.27]{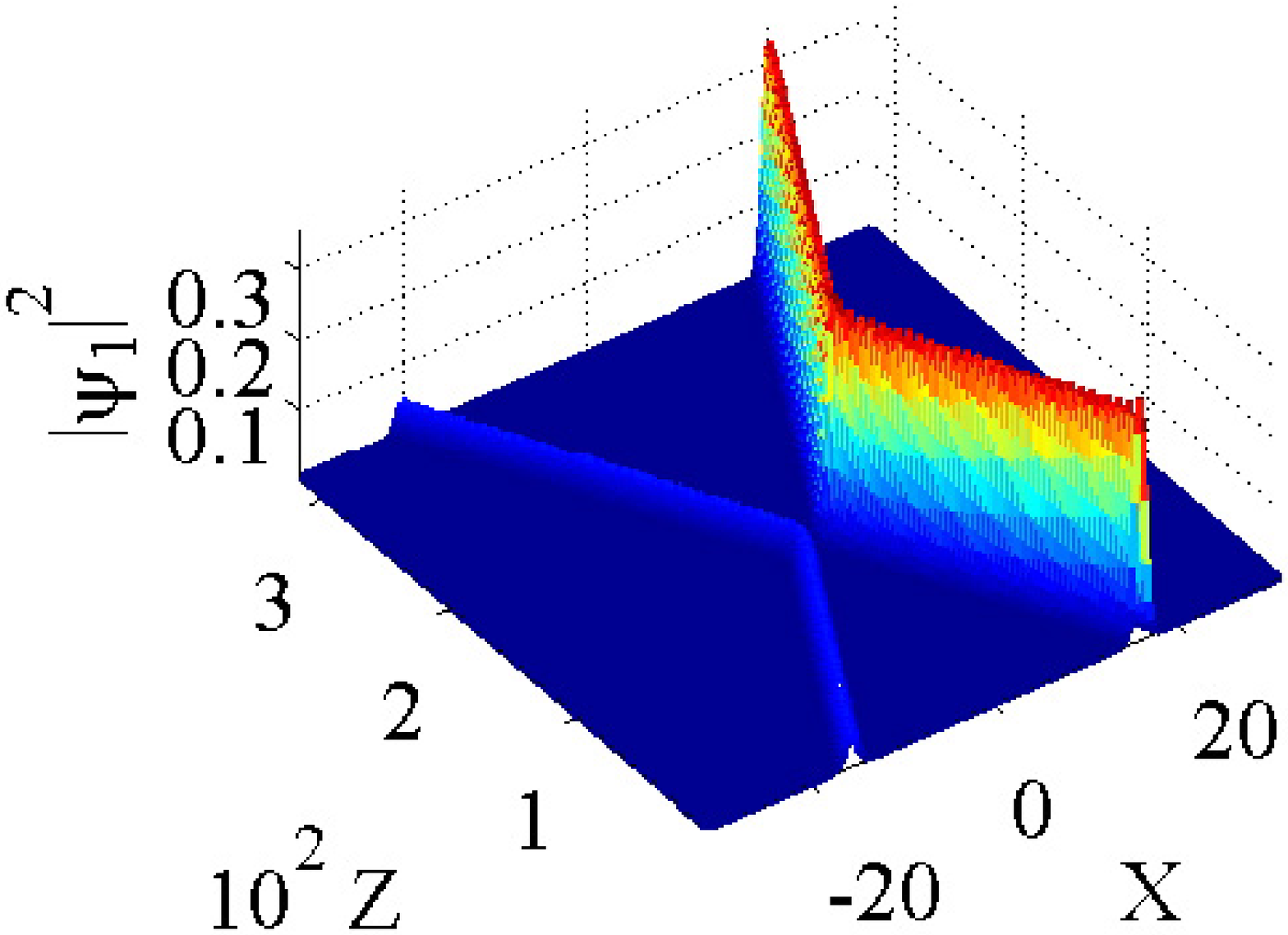}}%
\subfigure[] {\label{fig9b}
\includegraphics[scale=0.27]{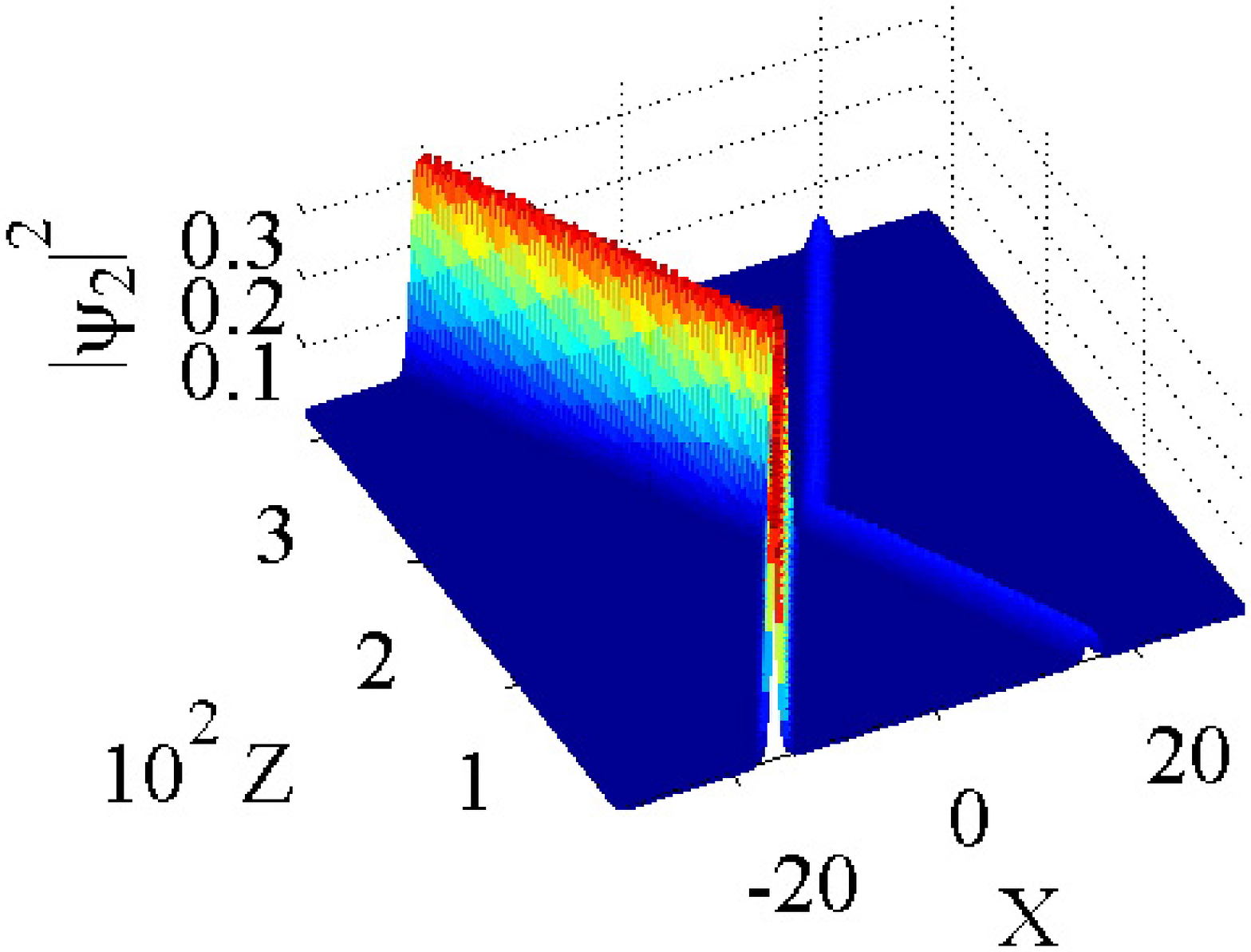}}
\subfigure[] {\label{fig9c}
\includegraphics[scale=0.28]{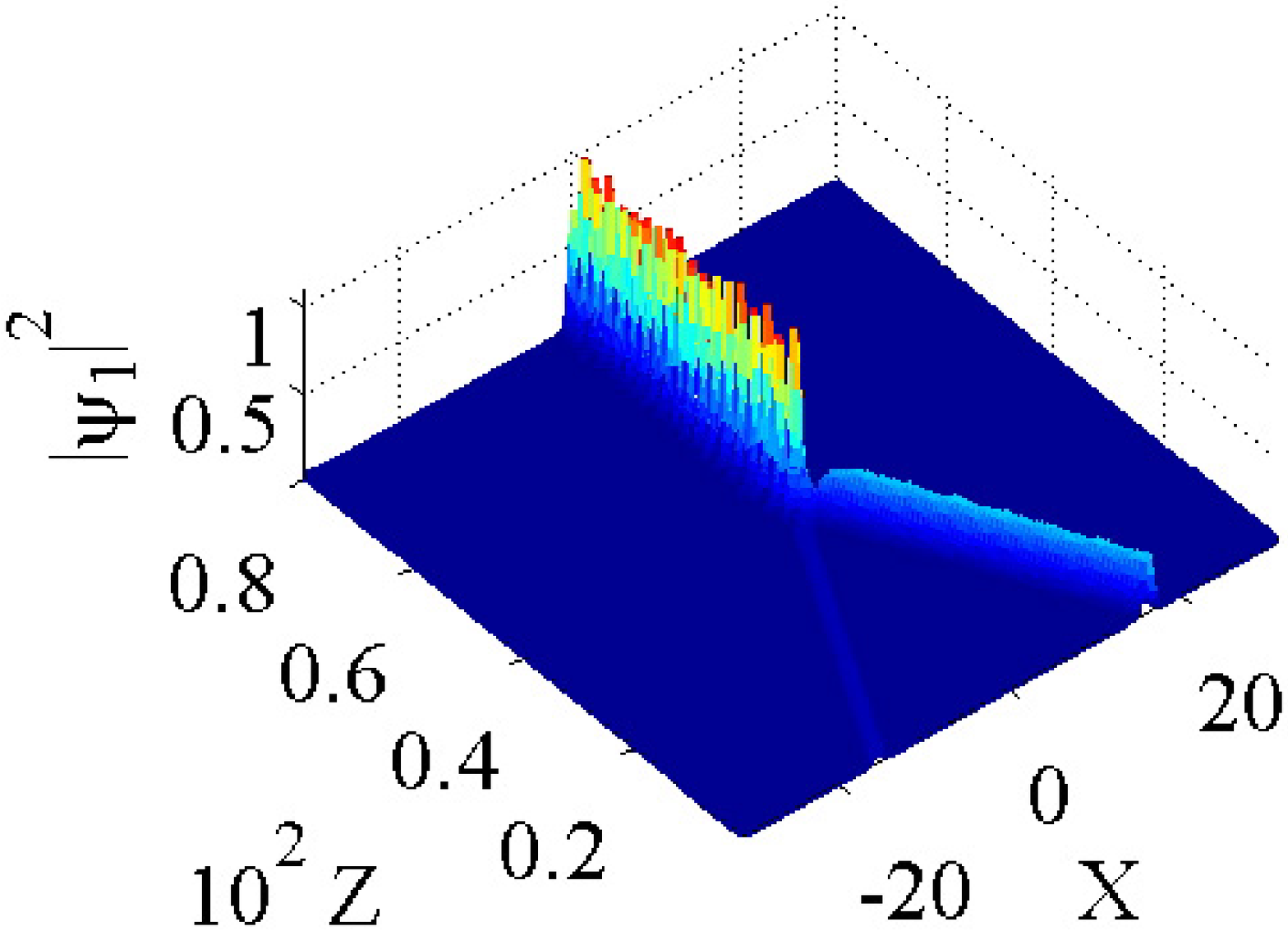}}
\subfigure[] {\label{fig9d}
\includegraphics[scale=0.28]{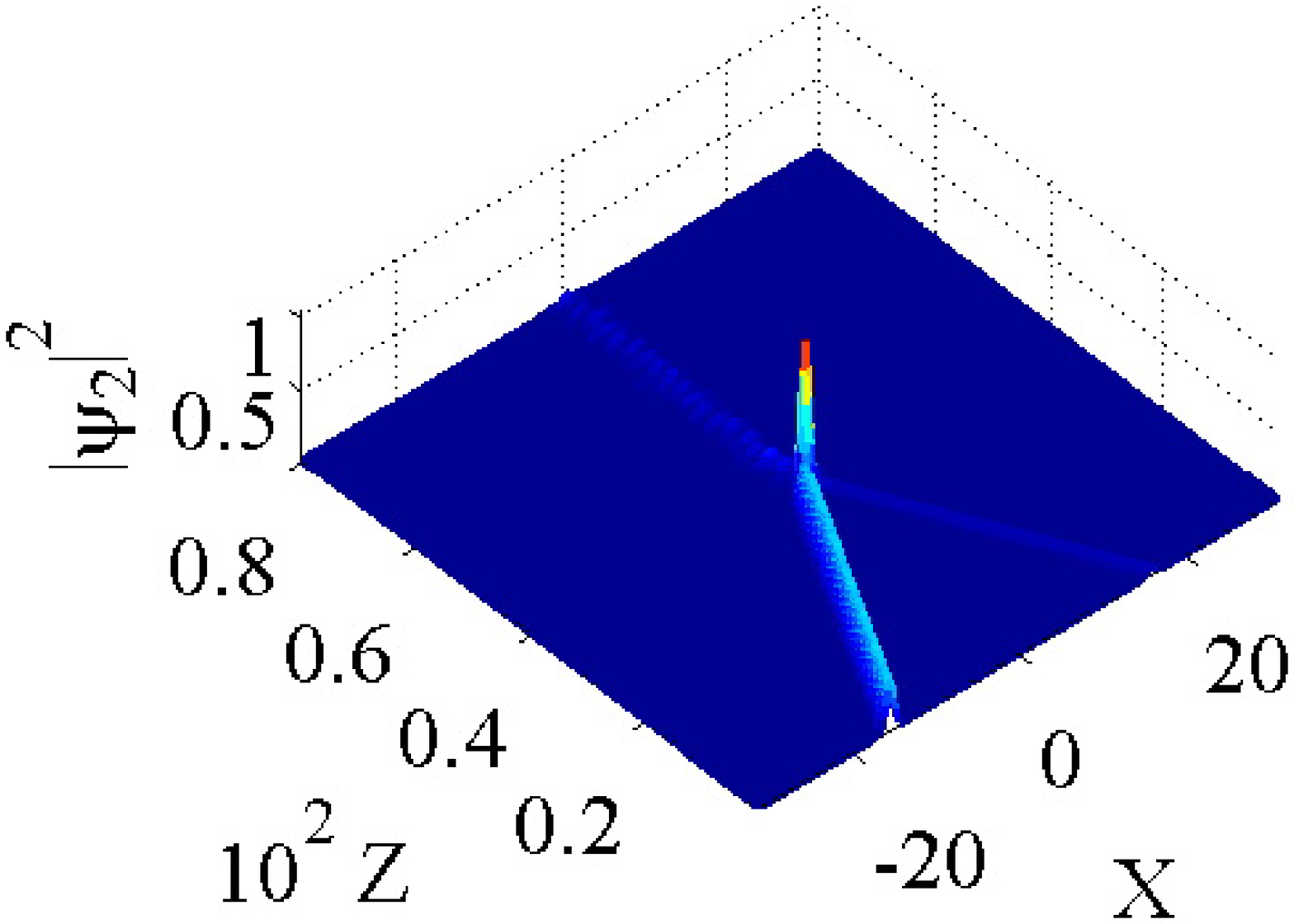}}%
\subfigure[] {\label{fig9e}
\includegraphics[scale=0.28]{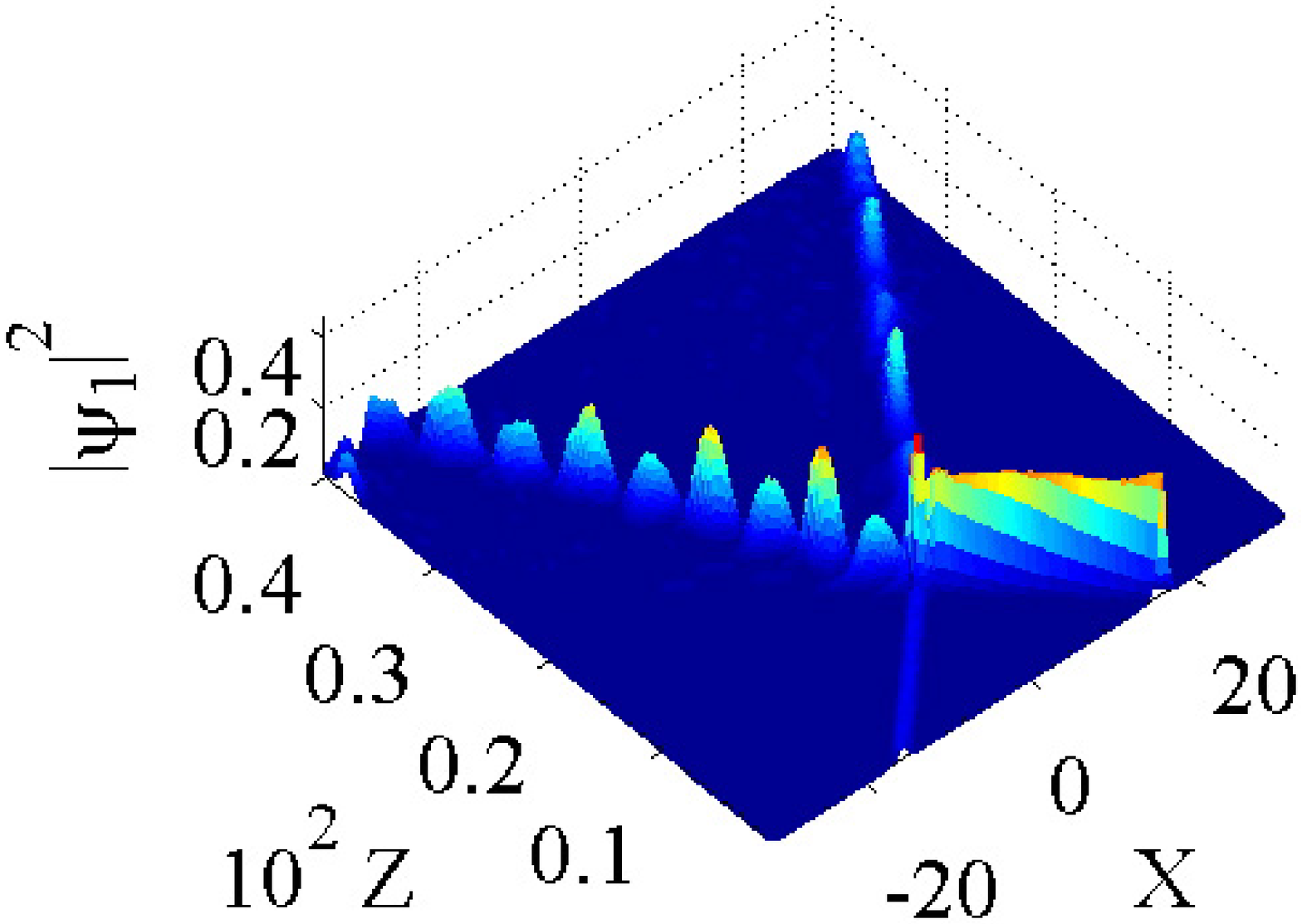}}
\subfigure[] {\label{fig9f}
\includegraphics[scale=0.28]{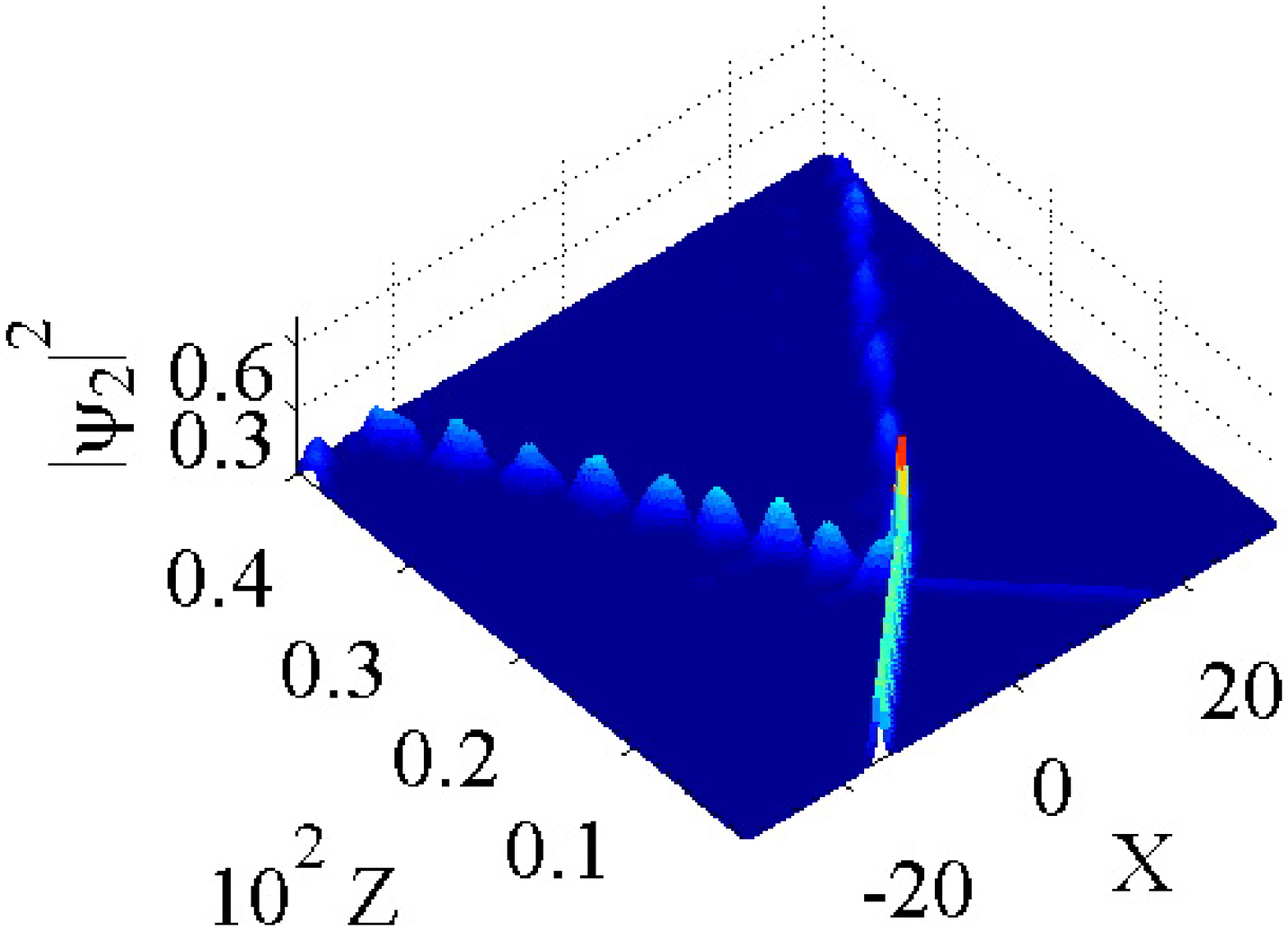}}
\caption{(Color online) Examples of collisions between asymmetric solitons
with opposite polarities: (a,b) for the slow solitons, with $\protect\eta %
=0.1$; (c,d) for the intermediate velocity, $\protect\eta =0.4$; (e,f) for
the fast solitons, $\protect\eta =1.6$.}
\label{Colli2}
\end{figure}

In Figs. \ref{Colli1} and \ref{Colli2},\ we display typical results of
collisions for asymmetric soliton pairs in the case which is close to the
border between the weakly- and strongly-bound systems, $(a,b,P)=(1,0.4,0.5)$%
. In panels \ref{Colli1}(a,b) and \ref{Colli2}(a,b), it is observed that
slowly moving solitons bounce back from each other elastically. When the
kick and ensuing velocities are larger, the collision becomes inelastic,
leading to the merger of the solitons into a single asymmetric breather.
Panels \ref{Colli1}(c,d) and \ref{Colli1}(e,f) demonstrate, severally, that
the merger of the colliding unipolar solitons may switch the polarity of the
emerging breather, in comparison with the original solitons, or produce a
breather which keeps the original polarity. The merger of the solitons with
opposite polarities gives rise to an asymmetric breather, whose polarity is
established spontaneously, as seen in Figs. \ref{Colli2}(c,d). Finally, at
still larger values of the kick, the moving solitons pass through each
other, reappearing in an excited form (each one as a moving breather), see
Figs. \ref{Colli1}(g,h) and \ref{Colli2}(e,f). The norms of the outgoing
vibrating solitons are equal, as they were before the collision.

The same sequence of outcomes of the collisions---rebound, merger into a
breather, and passage in the form of moving breathers---is observed, with
the increase of the kick ($\eta $), at other values of the parameters.
Collisions between symmetric solitons seem simpler (not shown here in
detail): Rebound at small values of $\eta $, and passage, without
conspicuous excitation of intrinsic vibrations, at larger $\eta $.


\section{The PT-symmetric version of the weakly coupled system}

It was proposed to realize $\mathcal{PT}$-symmetric systems in BEC by
linearly coupling two traps (cores, in the present terms) with the loss of
atoms in one of them, and compensating supply in the other, which may be
provided by a matter-wave laser \cite{PT-BEC}. The objective of the analysis
was to provide for a direct realization of the $\mathcal{PT}$ symmetry in
quantum media, after it was proposed \cite{PT-optics} and implemented \cite%
{Guo} in classical optics, see also review \cite{doulides}.

Accordingly, the $\mathcal{PT}$-balanced version of linearly-coupled GPEs (%
\ref{NLS}) is
\begin{gather}
i{\frac{\partial \psi _{1}}{\partial t}}=-{\frac{1}{2}}{\frac{\partial
^{2}\psi _{1}}{\partial x^{2}}}-\psi _{2}+i\gamma \psi _{1}  \notag \\
-\psi _{1}\int_{-\infty }^{+\infty }\left[ {\frac{|\psi _{1}(x^{\prime
})|^{2}}{(b^{2}+|x-x^{\prime }|^{2})^{3/2}}}-\frac{1}{2}(1-3\cos ^{2}\theta )%
{\frac{|\psi _{2}(x^{\prime })|^{2}}{(a^{2}+|x-x^{\prime }|^{2})^{3/2}}}%
\right] dx^{\prime },  \notag \\
i{\frac{\partial \psi _{2}}{\partial t}}=-{\frac{1}{2}}{\frac{\partial
^{2}\psi _{2}}{\partial x^{2}}}-\psi _{1}-i\gamma \psi _{2}  \notag \\
-\psi _{2}\int_{-\infty }^{+\infty }\left[ {\frac{|\psi _{2}(x^{\prime
})|^{2}}{(b^{2}+|x-x^{\prime }|^{2})^{3/2}}}-\frac{1}{2}(1-3\cos ^{2}\theta )%
{\frac{|\psi _{1}(x^{\prime })|^{2}}{(a^{2}+|x-x^{\prime }|^{2})^{3/2}}}%
\right] dx^{\prime },  \label{PTNLS}
\end{gather}%
where $\gamma >0$ is the coefficient of the gain and loss in the first and
second cores, respectively, $\kappa =G_{\mathrm{DD}}\equiv 1$ is fixed, as
above, and the local nonlinearity is dropped ($g=0$). As well as in the
recently studied model of the $\mathcal{PT}$-symmetric coupler with the
cubic nonlinearity \cite{PT,PT2}, stationary $\mathcal{PT}$-symmetric and
antisymmetric solutions to Eq. (\ref{PTNLS}) can be found as
\begin{eqnarray}
\psi _{1,2}^{\mathrm{(symm)}}(x,t) &=&e^{-i\mu t}\phi (x)\exp \left( \pm
\frac{1}{2}i\arcsin \gamma \right) \equiv e^{-i\mu t}\phi _{1,2}(x),
\label{symm} \\
\psi _{1,2}^{\mathrm{(anti)}}(x,t) &=&\pm ie^{-i\mu t}\phi (x)\exp \left(
\mp \frac{1}{2}i\arcsin \gamma \right) ,  \label{antisymm}
\end{eqnarray}%
where the upper and lower signs correspond to subscripts $1$ and $2$,
respectively, $\mu $ is a real chemical potential, and real function $\phi
(x)$ is the solution of stationary equation (\ref{phi}) for symmetric
solitons in the system without the $\mathcal{PT}$ terms, and with the same
value of $\mu $. Obviously, the symmetric and antisymmetric solitons exist
for $\gamma <1$, and they form continuous families parameterized by $\mu $,
like their counterparts in the conservative system.

\begin{figure}[tbp]
\centering%
\subfigure[] {\label{fig10a}
\includegraphics[scale=0.25]{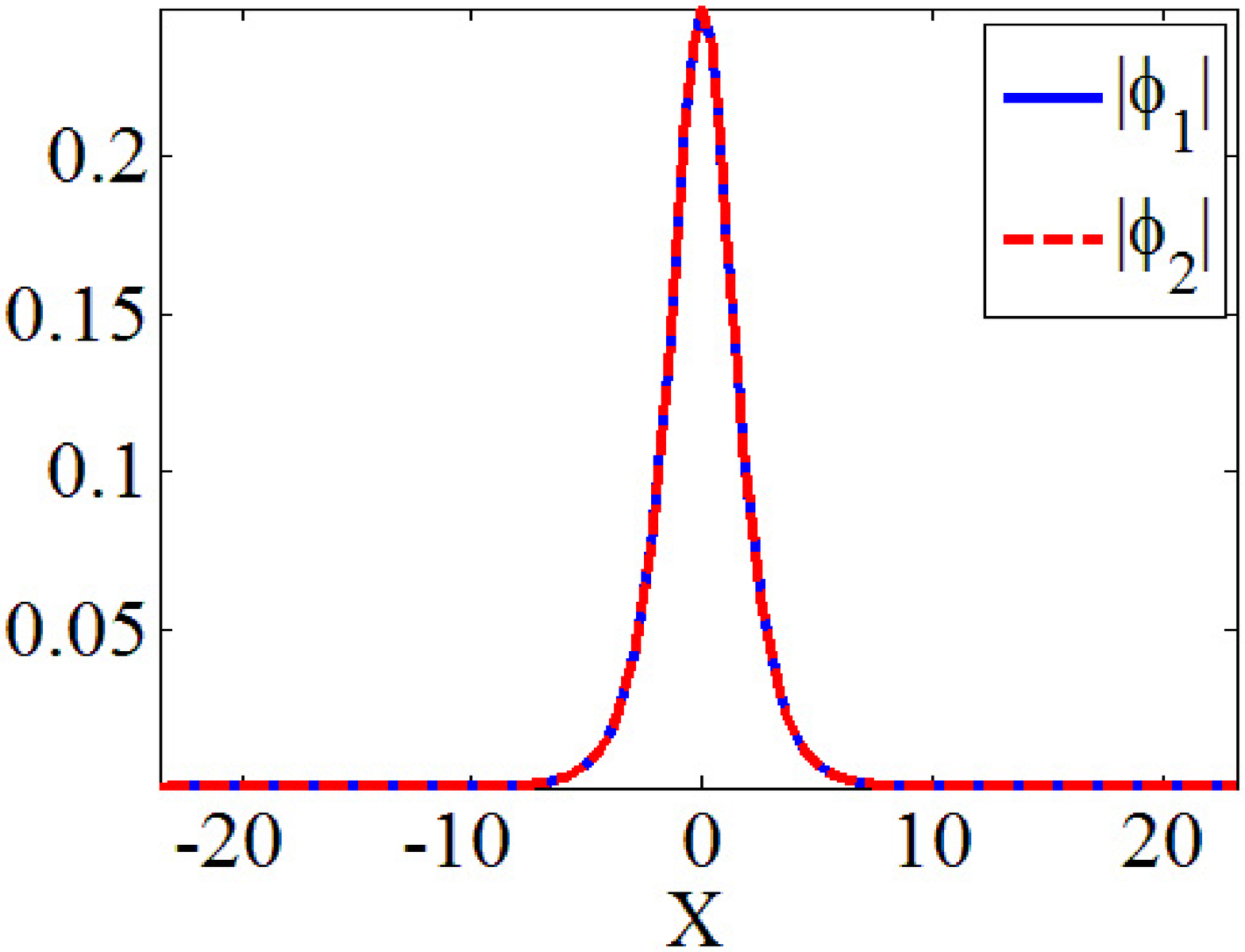}}%
\subfigure[] {\label{fig10b}
\includegraphics[scale=0.25]{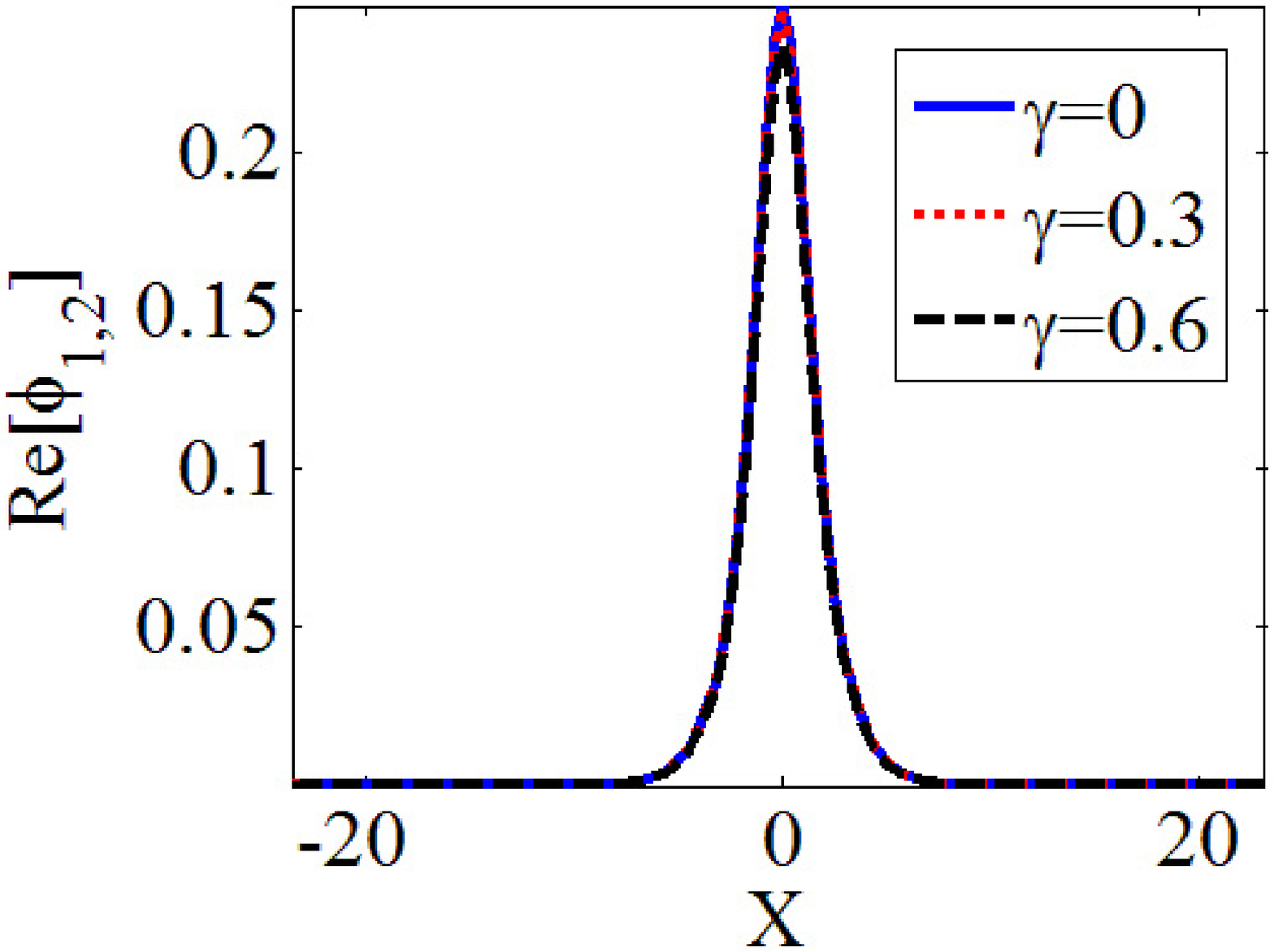}}
\subfigure[] {\label{fig10e}
\includegraphics[scale=0.25]{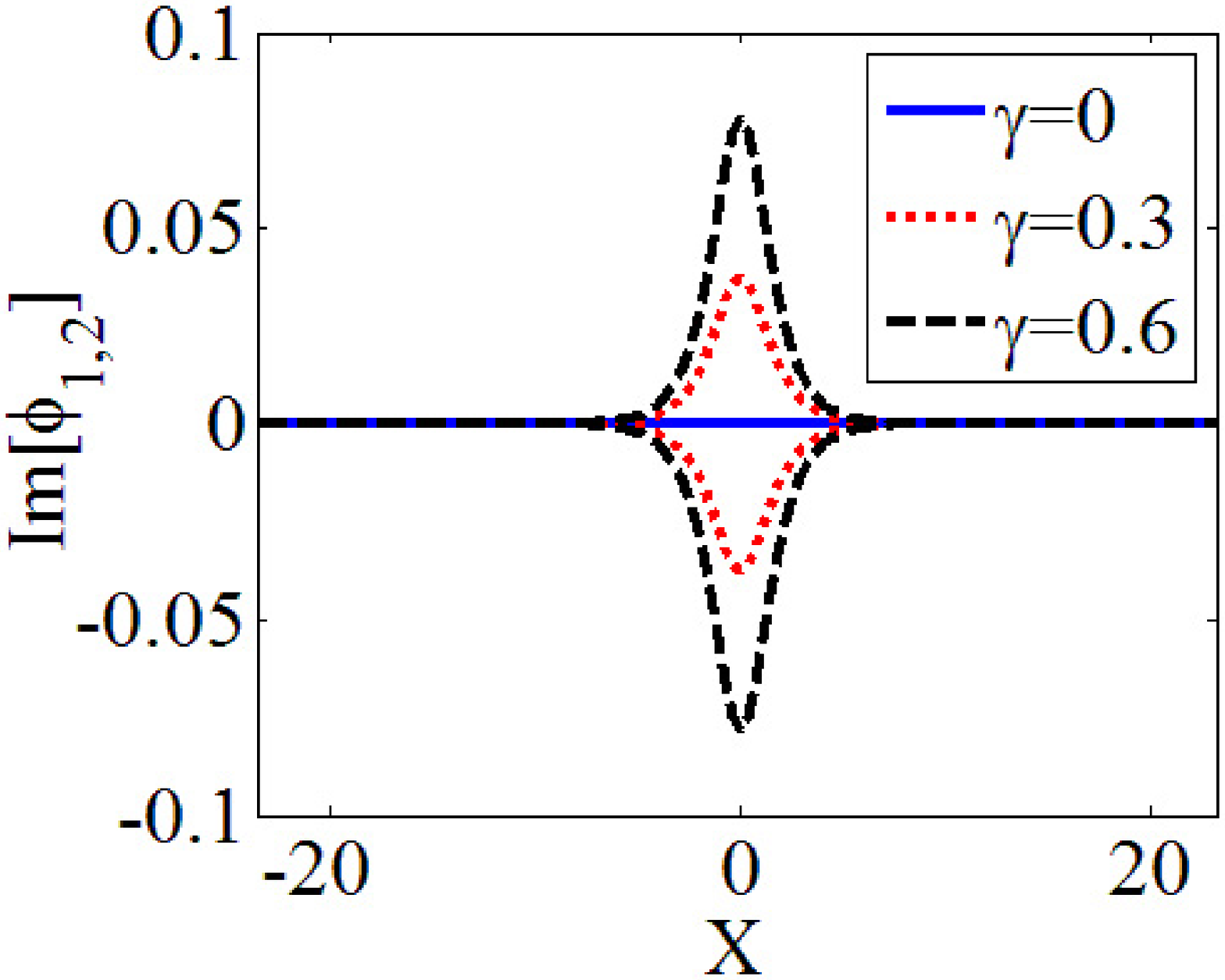}}
\caption{(Color online) Examples of stable $\mathcal{PT}$-symmetric
solitons, defined as per Eq. (\protect\ref{symm}), with $(a,b,P)=(1,0.4,0.3)$%
, for $\protect\gamma =0$, $0.3$, and $0.6$. (a) The profile of $\left\vert
\protect\psi _{1,2}(x)\right\vert $, which is common for the three solitons.
(b,c) Real and imaginary part of the solitons.}
\label{stablePT}
\end{figure}

Here we focus on the analysis of the stability of the $\mathcal{PT}$%
-symmetric and antisymmetric solitons (\ref{symm}), which is a nontrivial
problem in the present context. Typical examples of stable symmetric
solitons are displayed in Fig. \ref{stablePT}, along with their counterpart
in the system without the $\mathcal{PT}$ terms ($\gamma =0$).

\begin{figure}[tbp]
\centering%
\subfigure[] {\label{fig11a}
\includegraphics[scale=0.25]{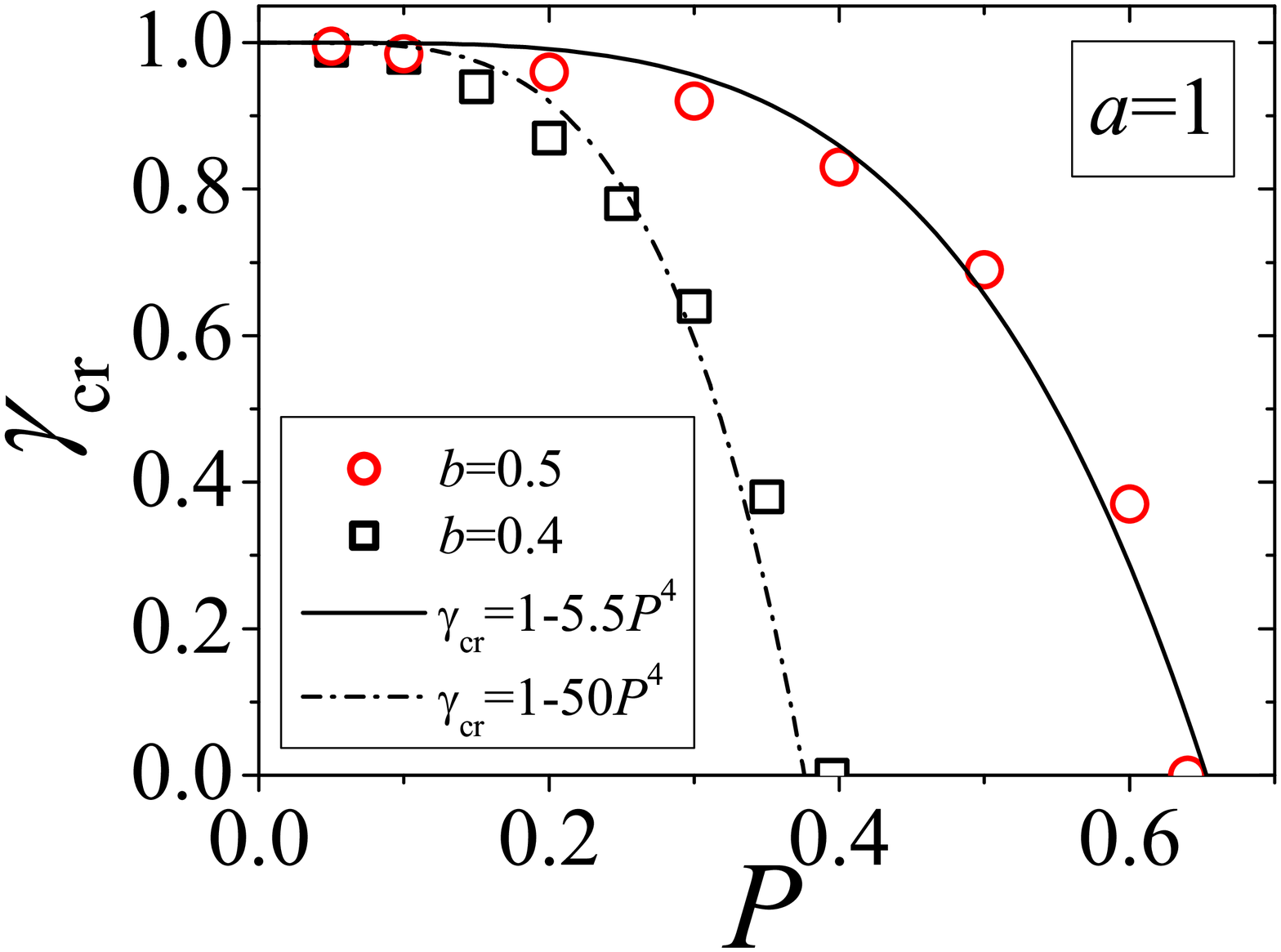}}%
\subfigure[] {\label{fig11b}
\includegraphics[scale=0.25]{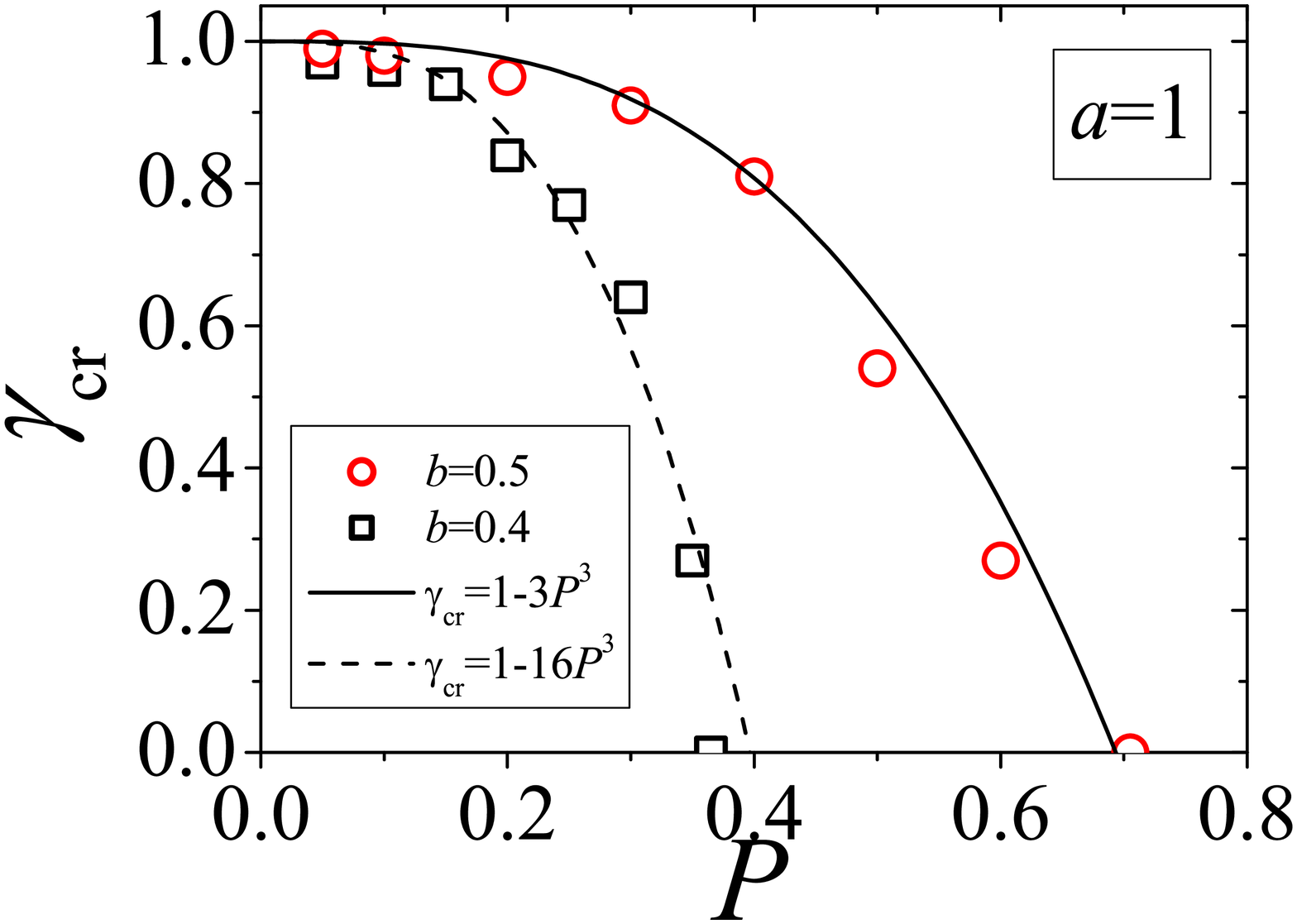}}
\caption{(Color online) Critical value $\protect\gamma _{\mathrm{cr}}$ (the
stability boundary for the $\mathcal{PT}$-symmetric solitons) versus the
total norm, $P$, at different fixed values of the pipe's parameter, $b$, in
the weakly-coupled system. (a) For the symmetric solitons (\protect\ref{symm}%
); (b) for antisymmetric ones (\protect\ref{antisymm}). The continuous and
dashed curves are guides for eye.}
\label{thresholdPT}
\end{figure}

The $\mathcal{PT}$-symmetric and antisymmetric solitons remain stable up to
a certain critical value, $\gamma _{\mathrm{cr}}$, of the gain-loss
coefficient, and are unstable in the interval of $\gamma _{\mathrm{cr}%
}<\gamma <1$, which is a situation typical for solitons in $\mathcal{PT}$%
-symmetric systems \cite{PT-soliton,PT,PT2}. The unstable solitons suffer a
blowup of the pumped component and decay of the damped one, which is a
typical scenario too (not shown here in detail). The most essential results
are presented in Fig. \ref{thresholdPT} for the weakly-coupled system, in
the form of the dependence of $\gamma _{\mathrm{cr}}$ on the total norm for
different diameters of the parallel-coupled pipes, $b$ (recall $b$ was
demonstrated above to be the most essential coefficient for the
weakly-coupled system). The figure demonstrates that the instability region
expands as the nonlinear interactions get stronger, which is caused either
by the increase of the total norm ($P$), or decrease of $b$.

\section{Conclusion}

The objective of this work is to extend the study of the symmetry breaking
of solitons in dual-core systems with the cubic nonlinearity and linear
coupling between the cores, that has been previously analyzed in full detail
for local interactions, to dipolar BEC with the long-range interactions,
which act both inside each core and between them. Two versions of the system
were introduced, weakly- and strongly-coupled ones, depending on the
relation between the diameter of the pipe-shaped traps and the distance
between them. In either case, the linear coupling accounts for hopping of
atoms between the cores. The symmetry-breaking bifurcation and stability
regions for symmetric and asymmetric solitons, as well as for
non-bifurcating antisymmetric ones, have been identified in both systems. In
addition to the solitons, the strongly-coupled system supports a stability
region for flat states with the unbroken symmetry between the cores, due to
the competition between the attractive and repulsive intra- and inter-core
dipole-dipole interactions. Collisions between kicked asymmetric solitons in
the weakly-coupled system were systematically studied too, showing bouncing
back at small velocities, merger into an asymmetric breather in the
intermediate range, and reappearance of vibrating fast solitons after the
collision. Finally, a $\mathcal{PT}$-symmetric generalization of the weakly
coupled system was introduced as a more ``exotic" extension of the system,
and a stability boundary for $\mathcal{PT}$-symmetric and antisymmetric
solitons was found.

A challenging generalization of the analysis may be its application to 2D
dual-core systems, where, in particular, the symmetry breaking may occur not
only for fundamental solitons, but also for solitary vortices.

\begin{acknowledgments}
We appreciate a valuable discussion with L. Santos. This work was supported
by Chinese agency CNNSF (grants No. 11104083, 11204089, 11205063), by the Guangdong Provincial Science and technology projects (2011B090400325), by the German-Israel Foundation through grant No. I-1024-2.7/2009, and by the Tel
Aviv University in the framework of the \textquotedblleft matching" scheme
for a postdoctoral fellowship of Y.L.
\end{acknowledgments}

\bibliographystyle{plain}
\bibliography{apssamp}

\end{document}